\documentclass[preprint2]{aastex}

\newcommand{\etal}{et~al.}

\newcommand{\Msun}{M_{\odot}}

\newcommand{\KE}{E_{\rm K}}
\def\gsim{\mathrel{\rlap{\lower 4pt \hbox{\hskip 1pt $\sim$}}\raise 1pt
\hbox {$>$}}}
\def\lsim{\mathrel{\rlap{\lower 4pt \hbox{\hskip 1pt $\sim$}}\raise 1pt
\hbox {$<$}}}

\begin{document}

\title{Nebular Spectra of SN 1998bw Revisited: \\
Detailed Study by One and Two Dimensional Models}

\author{
K.~Maeda\altaffilmark{1}, 
K.~Nomoto\altaffilmark{2,3}, 
P.A.~Mazzali\altaffilmark{3,4,5},
J.~Deng\altaffilmark{6,2}, 
}

\altaffiltext{1}{Department of Earth Science and Astronomy,
Graduate School of Arts and Science, University of Tokyo, Meguro-ku, Tokyo
153-8902, Japan: maeda@esa.c.u-tokyo.ac.jp}
\altaffiltext{2}{Department of Astronomy, School of Science,
University of Tokyo, Bunkyo-ku, Tokyo 113-0033, Japan}
\altaffiltext{3}{Research Center for the Early Universe, School of
Science, University of Tokyo, Bunkyo-ku, Tokyo 113-0033, Japan}
\altaffiltext{4}{Instituto Nazionale di Astrofisica 
(INAF)-Osservatorio Astronomico di Trieste, Via Tiepolo 11, 
I-34131 Trieste, Italy}
\altaffiltext{5}{Max-Planck-Institut f\"ur Astrophysik, 
Karl-Schwarzschild-Stra{\ss}e 1, 85741 Garching, Germany}
\altaffiltext{6}{National Astronomical Observatories, CAS, 
20A Datun Road, Chaoyang District, Beijing 100012} 

\begin{abstract}
Refined one- and two-dimensional models for the 
nebular spectra of the hyper-energetic Type Ic supernova (SN) 1998bw, 
associated with the gamma-ray burst GRB980425, from 125 to 376 days 
after B-band maximum are presented.  One dimensional,
spherically symmetric spectrum synthesis calculations show that 
reproducing features in the observed spectra, i.e., 
the sharply peaked [OI] 6300\AA\ doublet and MgI] 4570\AA\
emission, and the broad [FeII] blend around 5200\AA, requires the existence of a
high-density O-rich core expanding at low velocities ($\lsim 8,000$ km
s$^{-1}$) and of Fe-rich material moving faster than the O-rich material. 
Synthetic spectra at late phases from aspherical (bipolar) explosion models 
are also computed with a two-dimensional spectrum synthesis code. 
The above features are naturally explained by
the aspherical model if the explosion is viewed from a direction close to the
axis of symmetry ($\sim 30^{\rm o}$), since the aspherical model yields a
high-density O-rich region confined along the equatorial axis.  By examining a
large parameter space (in energy and mass), our best model gives following physical 
quantities: the kinetic energy $E_{51} \equiv E_{\rm
K}/10^{51}$ ergs $\gsim 8 - 12$  and the main-sequence mass of the progenitor
star $M_{\rm ms} \gsim 30 - 35 \Msun$. The temporal spectral evolution of SN
1998bw also indicates mixing among Fe-, O-, and C-rich regions, 
and highly clumpy structure. 
\end{abstract}


\section{INTRODUCTION}

There has been accumulating evidence that a class of supernovae (SNe) is
related to long-duration  Gamma-Ray Bursts (GRBs; see Piran 1999 for a review)
and possibly to their low energy analog, X-Ray Flashes (XRFs; Heise et al. 2001).  
The discovery of SN 1998bw in the error box of GRB 980425 
(Galama et al. 1998) raised the issue, and now the association between GRBs and SNe is 
firmly confirmed by the emergence of supernova
spectra in the optical afterglows of some GRBs, i.e., SN 2003dh/GRB 030329  (Hjorth et al. 2003;
Kawabata et al. 2003; Matheson et al. 2003; Stanek et al. 2003), SN 2003lw/GRB 
031203 (Malesani et al. 2004; Thomsen et al. 2004, Mazzali et al., in
preparation), and SN 2002lt/GRB 021211 (Della Valle et al. 2003). 

All these supernovae (except SN 2002lt) seem to belong to a special class (see
e.g., Matheson 2004 for a review). 
Their spectra near maximum optical brightness are  characterized
by significant blending of  very broad lines.  A popular idea to explain this
feature in the GRB-related supernovae is a hyper-energetic explosion ($E_{51}
\equiv  E_{\rm K}/10^{51}$ ergs $\gsim 5 - 10$) of a very massive star ($M_{\rm
ms} \gsim 20 - 25\Msun$).  Such an energetic supernova is often called a 
"hypernova" (Iwamoto et al. 1998; Nomoto et al. 2004), with a somewhat
different use of the original terminology suggested by Paczynski (1998) to
describe the entire GRB/afterglow phenomenon. 

The nature of GRBs and hypernovae, and their mutual relation, are still the
subject of debate.  Asphericity seems to be the
key to understand both GRBs and hypernovae.  It is widely accepted that GRBs
are produced by a relativistic jet viewed on or near its axis (Frail et al.
2001; Bloom, Frail, \& Kulkarni 2003).  Their physical link to hypernovae then
suggests that hypernovae also should show signatures of asphericity, 
according to popular progenitor scenarios: 
a black-hole plus an accretion disk (Woosley 1993; MacFadyen
\& Woosley 1999; Brown et al. 2000), or a highly magnetized neutron star
(Nakamura 1998; Wheeler et al. 2000). 

The hypernova models for the GRB-related supernovae have been developed on the
basis of modeling early phase observations up to $\sim$ 2 months after the
explosion.  For SN 1998bw, photometric and spectroscopic observations covering
more than 1 year are available (Sollerman et al. 2000; Patat et al. 2001). 
Since SN 1998bw was the first supernova being identified as the counterpart of
a GRB (GRB980425), and it has been extensively referred to as a prototypical 
hypernova, it is important to understand its nature. 

Studying the late phase light curve and spectra has provided additional hints
on the nature of SN 1998bw.  While the early phase observations can be
reproduced by a "spherical" hypernova model  (Iwamoto et al. 1998; Woosley,
Eastman, \& Schmidt 1999; Nakamura et al. 2001), the light curve after $\sim 2$
months declined less steeply than the spherical hypernova model prediction 
(McKenzie \& Schaefer 1999).  A similar problem exists for other hypernovae
(e.g., Mazzali, Iwamoto, \& Nomoto 2000; Yoshii et al. 2003) and some SNe
IIb/Ib/Ic (e.g., Clocchiatti \& Wheeler 1997). For hypernovae, Maeda et al.
(2003a) attributed the failure of the spherical models to the possible
existence of an inner low-velocity dense core, which may be formed as the
consequence of an aspherical explosion  (e.g., Maeda \& Nomoto 2003b). 

Late phase spectra provide an excellent tool to examine the geometry of the
ejecta, as they can probe deeper than the early-phase spectra.  The nebular
spectra of SN 1998bw showed peculiar features which were not explained by a
simple picture (Patat et al. 2001).  First, the [OI] 6300\AA\ \& 6363\AA\
doublet shows a sharply peaked emission profile.  The profile is different from
both the parabolic profile expected from a homogeneous spherical distribution of
emitting material and the flat-topped boxy profile expected from material
distributed in a shell.  Sollerman et al. (2000) found that both the
spherical hypernova model CO138 of Iwamoto et al. (1998) 
(CO138E30; the explosion with $E_{51} = 30$ of a $13.8 \Msun$ CO star 
evolved from a zero-age main sequence star of $M_{\rm ms} \sim 40 \Msun$), 
and CO6C of Woosley et al. (1999) 
($E_{51} = 22$, a 6.55 $\Msun$ CO star, and $M_{\rm ms} \sim 25\Msun$) 
yielded too broad [OI] 6300\AA\ emission.  The feature near
5200\AA, interpreted as a blend of [FeII] in Mazzali et al. (2001) and Maeda et
al. (2002) also shows a peculiarity:  it is broader than the [OI] 6300\AA.
Mazzali et al. (2001) showed that each line contributing to the [FeII] blend
should be broader than the [OI] 6300\AA, i.e., velocities of once-ionized iron
ions should be higher than those of neutral oxygen along the line of sight. 
Subsequently, Maeda et al. (2002) computed the theoretical line profiles
expected from bipolar aspherical models.  They showed that the peculiarity of
the [OI] 6300\AA\ and [FeII] 5200\AA\ emissions in SN 1998bw can be explained
by the aspherical explosion models, if the axis of symmetry is roughly aligned
with the line of sight. 

Although these previous studies revealed the peculiarities and suggested 
possible solutions, there were still some limitations.  
As for one dimensional models,  previous studies (Sollerman et al. 2000;
Mazzali et al. 2001) did not model the  full set of late-phase spectra at
various epochs on the basis of a hydrodynamic model. 
Concerning two dimensional models, 
the analysis of Maeda et al. (2002) had the following limitations. 
(1) First, they applied a simplified time-independent computation for the line 
profiles. 
(2) They did not model the flux. 
A more detailed analysis  based on nebular line emission physics is necessary
to derive the nature, e.g., the explosion energy, of the explosion.  
(3) Also, it should be examined if their models, 
focusing on the [FeII] 5200\AA\ and the [OI] 6300\AA, are consistent with other
emission lines, e.g., MgI] 4570\AA\ and [CaII] 7300\AA.  (4) Their model grid
(in energy and mass) was rather coarse.  To constrain physical values such as
the energy, models with various energies and masses should be examined
comprehensively. 
(5) Finally, they only examined one nebular spectrum
(+216 days after the maximum optical luminosity). It is important to examine if
"one" model can consistently reproduce the temporal evolution of spectra.

The purpose of this paper is to perform a more detailed analysis of the nebular 
spectra of SN 1998bw than previous works. 
In addition to one dimensional spectrum synthesis, we use a two dimensional
nebular spectrum synthesis code, which includes two-dimensional $\gamma$-ray
transfer and a computation of ionization/thermal structures for
two-dimensional density and abundance distributions.  First, the
nebular spectra of SN 1998bw are examined by means of one dimensional spectrum
synthesis in \S 2, to obtain conditions necessary to explain the spectra of SN
1998bw.  The subsequent parts are devoted to the two dimensional models.  
In \S 3 we describe models and method, and present a comparison between 
the synthetic spectra and the observations. Summary and 
discussion are presented in \S 4.

\section{ONE DIMENSIONAL MODELS}

\subsection{Method}

We use a non-LTE code to compute nebular spectra.  The code is an extension of
the one-zone nebular code of Mazzali et al. (2001), allowing radially
stratified density and abundance distributions as an input model.  
First, a Monte-Carlo Method 
(e.g., Cappellaro et al. 1997; Chugai 2000; Maeda et al. 2003a) 
is used to solve the transport of $\gamma$-rays 
released by the decay chain $^{56}$Ni $\to$ $^{56}$Co$\to$ $^{56}$Fe.  
Positrons released by the same decay chain 
are assumed to be trapped {\it in situ} (Axelrod 1980). 
The light curve of SN 1998bw suggests that at a 
few hundred days $\gamma$-rays are still the dominant heating source  
(Nakamura et al. 2001; Maeda et al. 2003a), making the heating of the ejecta 
little sensitive to the detail of positron transport 
(e.g., Milne, The, \& Leising 2001). 

Next, ionization and NLTE thermal balance in each shell are solved 
on the basis of the prescription given by Ruiz-Lapuente \& Lucy (1992).  
For the ionization source, we consider only impact ionizations by high-energy
particles produced by the deposition of $\gamma$-rays (and $e^{+}$).  
The ionization balance is then solved by equating the impact ionization rate to the 
recombination rate for each ion. An assumption here is that photoionization is negligible. 
Although it is still a matter of debate (see e.g., Sollerman et al. 2004), it is argued 
that large multiple resonance scattering at high energies effectively reduces the UV 
radiation, therefore reduces the photoionizations (Kozma \& Fransson 1998b; Kozma et al. 
2005). 
Level populations are obtained by solving rate equations in steady state for each 
level along with the thermal balance (i.e., equating of the non-thermal heating 
rate and the line cooling rate). We consider mainly forbidden lines as a source of cooling 
(Ruiz-Lapuente \& Lucy 1992), but strong allowed transitions, e.g., CaII IR, are also 
included. Under nebular conditions, radiative losses are entirely due to collisionally 
excited lines. We do not take radiation transfer effects into account, since the 
epochs examined in the present work are later than $100$ days so that radiation transfer 
effects are usually small at least at optical wavelengths (see \S 4.5.1 for more 
detail). Observationally, after day $100 \sim 200$, neither the feature at 4570\AA\ 
nor that at 6300\AA\ evolved significantly (Patat et al. 2001), indicating that 
these features are totally dominated by forbidden lines and that radiation transfer 
effects are negligible. In particular, we do not include line scattering and fluorescence, 
an assumption that is usually made in nebular analysis (e.g., Kozma et al. 2005). 
Fluorescence may have effects in some lines, i.e., Ca II H, K and IR-triplet, and NaI 
(Kozma \& Fransson 1998b), for which we do not present detailed modeling in the 
present work. 
Finally, a synthetic spectrum is obtained 
by integrating the emission from each shell. 

Throughout the paper, synthetic spectra are compared with the spectra 
of SN 1998bw taken at 125, 200, 337, and 376 days after B maximum 
(Patat et al. 2001). The first spectrum ($+125^{\rm d}$) 
may not fully be nebular and hence can not be fully interpreted 
using the present computational code (see \S 4). 
Despite this, we include this spectrum in the comparison. 
The spectrum is at least partly nebular and therefore can still be used 
to reject any model producing a spectrum very different from it.

\subsection{SN 1998bw: One Dimensional Approach}

With the time-dependent computational code, we first examine the synthetic 
spectra for  the original spherical hypernova
model CO138E30 (Iwamoto et al. 1998; Nakamura et al. 2001).  
The mass cut, i.e., the boundary between the collapsing core and 
the ejecta, chosen in the present work is deeper than the original, 
yielding the mass of $^{56}$Ni $\sim 0.8 \Msun$ in order to roughly fit 
luminosities at the late phases 
(The original model with $^{56}$Ni $\sim 0.4 - 0.6 \Msun$ is fainter than 
the late-phase observations; Nakamura et al. 2001). 
Figure 1 shows the temporal evolution of the synthetic spectra as compared with
the observed ones.  
We next try to obtain better fits by changing the density
structure below $v \lsim 10,000$ km s$^{-1}$ and the abundance distribution
throughout the ejecta (described below  in more detail). 
The temporal sequence of the spectra of this
"modified" CO138E30 model is shown in Figure 2.  The density structure and
abundance distribution are shown in Figures 3 and 4, respectively. 

The "original" model has difficulties in reproducing 
both (a) line profiles in each single spectrum 
and (b) temporal evolution. 
As for the line profiles (a), elements in outer layers such as 
O and Mg produce synthetic lines
with very wide, flat-topped profiles, while the observed 
[OI] 6300\AA\ and MgI] 4570\AA\ are narrow and sharply peaked. 
The flat-topped profile
is the consequence of the distribution of emitting material with a central hole
(See Figures 3 and 4a: O is distributed at $\gsim 12,000$ km s$^{-1}$). 
Contrary to the observations, the synthetic [FeII] 5200\AA\ blend is narrower
than the synthetic [OI] 6300\AA. 
This is a typical character of any spherically
symmetric evolution-collapse-explosion model, as was pointed out by Mazzali et
al. (2001). 
The problem in the temporal evolution (b) is that 
the "original" CO138E30 model yields very rapidly fading lines [OI] 6300\AA\ 
and MgI] 4570\AA\ between the epochs $+125^{\rm d}$ and $+200^{\rm d}$ (Fig. 1). 

The modified model (Fig. 2) is constructed so as to overcome these problems. 
The line profiles suggest that Fe (mainly  synthesized as $^{56}$Ni
at the explosion) is on average distributed at higher velocities than hydrostatic burning
products, e.g., O and Mg. 
The sharply peaked [OI] and MgI] lines can be accounted for 
if these lines are predominantly emitted at low velocity. 
In the one dimensional
representation, a peculiar abundance distribution is therefore required:  A
high density O-rich core is added to the original CO138E30 model.  
It also helps to explain the more
rapid decline of the [FeII] than the [OI] 6300\AA\ and the MgI] 4570\AA\ 
emission (Patat et al. 2001), since the contribution of higher velocity
material becomes smaller as time goes by. 
Because deposition rate is larger in the modified model than in the original one, 
the mass of $^{56}$Ni is smaller in the former, $M$($^{56}$Ni) $\sim 0.6 \Msun$. 

Another feature of the modified model is some mixing between the Fe-rich layer
(originally at velocities at $\sim 8,000 - 12,000$ km s$^{-1}$) 
and the O-rich core (added at $\lsim 8,000$ km s$^{-1}$).  
The observed [FeII] 5200\AA\ and [CaII] 7300\AA\ lines show mildly 
peaked profiles (though less sharply peaked than the [OI] and MgI]), 
suggesting that a small fraction of the Fe-rich materials are mixed downward 
into the low velocity O-rich core. 
The suggestion that (at least in a spherically symmetric representation) 
mixing is necessary was already made by Sollerman et al. (2000). Our model 
represents an even more unusual structure (from the point of view of 1D hydrodynamics), 
since the O-rich core is located at low velocities together with just a small 
fraction of Fe. This helps to keep the ionization of Fe as high as what is 
inferred from the observed spectra, although the usual mixing (Sollerman et 
al. 2000) with an Fe-dominated central region will lead to too low an ionization.

The mixing is also favored in view of the temporal evolution 
of each line intensity. 
Without any mixing, energy deposited by $\gamma$-rays and positrons into 
the O-rich region is all reprocessed into oxygen lines, 
predominantly into the [OI] 6300, 6363\AA\ doublet. 
Therefore, the temporal evolution of [OI] 6300\AA\ emission 
roughly traces the deposition rate, yielding too rapid fading of the line 
from $+125^{\rm d}$ to $+200^{\rm d}$ (Fig. 1). 
If other elements such as iron are mixed in, then the situation is 
different. Intensity of each line depends on contribution from others, which 
also depends on thermal condition within the nebula. 
In the materials composed with both iron and oxygen, 
the intensity ratio [FeII] 5200\AA/[OI] 6300\AA\ decreases rapidly with
decreasing density, yielding larger relative intensity of [OI] 6300\AA\ in 
more advanced epochs. 
Given that the sum of the intensities of these lines roughly follows 
the deposition rate, increasing relative importance of [OI] 6300\AA\ with time 
yields the [OI] less rapidly fading than in the absence of any mixing. 

The modified model (Figs. 3 and 4b) is constructed by mixing Fe from the
Fe-rich layer downward into the O-rich core and compensate for it by mixing
oxygen from the O-rich core into the Fe-rich layer.  The same is also done for
Ca.  In addition, intensity of the [OI] 6300\AA\ at the first epoch is further
decreased by introducing clumping in the O-rich region. 
Regarding clumping, first introduced for SN 1998bw by Mazzali et al. (2001), the
following facts give hints: (1) Without clumping [FeIII] is too strong at $\sim
4700$\AA, which is not seen in the observations. (2) The intensity ratio of
[CaII] 7300\AA\ to CaII IR is too large in the original CO138E30 model. 
Therefore, we have introduced a filling factor of 0.1 throughout the ejecta
The filling factor is kept unchanged at all epochs.

Comparing Figures 1 and 2, it is seen that the 
fit to the observed spectral sequence is very much improved.  
There are still some difficulty in reproducing the spectrum at $+125^{\rm d}$ 
(while it is much better than the original model). 
In addition to the strong [OI] 6300\AA\ emission in the synthetic spectrum, 
MgI] 4570\AA\ was also too strong 
at the first epoch (In Fig. 2, the mass fraction of Mg is reduced by a factor of 10 
at the first epoch relative to the subsequent epochs). 
However, it may simply be that the density in the first epoch is 
too high to apply the nebular 
spectral computation (see \S 4). 
The present model is in a sense {\it ad hoc}, introducing peculiar elements distribution 
and density structure (i.e., O-rich core). We will show in \S 3 that 
this apparent peculiarity in one-dimensional representation 
can naturally be interpreted in the context of a two-dimensional model.

\section{TWO DIMENSIONAL MODELS}

\subsection{Method and Models}

We have developed a two-dimensional code, which is applicable 
to any two-dimensional distribution of density
and abundances (including the distribution of the heating source 
$^{56}$Ni). In the current version, axisymmetry along the $z$-axis (polar axis)
and reflection symmetry on the equatorial plane are assumed. 
The included physics is the same with the one dimensional code. 
The difference between the one- and two-dimensional versions 
is in the treatment of the $\gamma$-ray deposition and the
computation of the whole spectrum. 
The $\gamma$-ray deposition is solved as a two dimensional radiation transport
problem using the Monte-Carlo Method.  
Ionization and NLTE thermal balance are solved in each mesh zone independently, 
since these processes take place locally as long as the optical depth is 
negligibly small. 
A synthetic spectrum is then computed for different orientations 
(divided uniformly into 10 degree angular zones), 
taking into account different Doppler shifts for different orientations. 

We use as input the models presented in Maeda et al. (2002). These include
spherical and aspherical explosions of a $16 \Msun$ He star, the core of a star
of $M_{\rm ms} = 40 \Msun$.  Asphericity was generated assuming angle-dependent
energy deposition, preferentially concentrated toward the polar direction, 
at the center of the collapsing core (See Maeda et al. (2002) for details). 

The models we explore in this paper are listed in Table 1.  The distributions
of density and of a few selected elements in the homologous expansion phase are
presented in Figure 5. 
We examine models with various degree of asphericity 
defined by the parameter BP. The value of BP is the axial ratio of the initial 
aspherical energy injection. See Maeda et al. (2002) for details (their parameter 
$\alpha/\beta$ is the same as BP). 
The models correspond to those of Maeda et al. (2002)
as follows:  BP16 (A), BP8 (C), BP2 (E), and a spherical model BP1 (F).  In the
present work, we add an additional model (BP4) which has an asphericity
intermediate between models BP8 and BP2, by repeating the calculations of Maeda
et al. (2002) but with a different asphericity parameter. 

As seen in Table 1, we generate a series of models with various kinetic energy 
by artificially multiplying the velocities at any points in the ejecta 
by a factor $f_V$. 
The factor $f_V$ is taken as a parameter, ranging from 0.7 to 1.9. 
For each model specified by the parameters BP and $f_V$, 
the masses of oxygen ($M_{\rm O}$) and $^{56}$Ni in the ejecta are changed by hand 
in order to obtain a good fit to the intensities of the [OI] 6300\AA, 6363\AA\
and the [FeII] 5200\AA\ (see \S 3.2 for details).  
The kinetic energy in the ejecta is scaled as 
$E_{\rm K} \propto M_{\rm ej} {f_V}^2$, 
where $M_{\rm ej}$ is the ejecta mass, mainly oxygen, 
obtained by spectral fitting. 
Because of the procedure, 
the oxygen mass and therefore the kinetic energy are different from 
the original values ($M_{\rm O} \sim 8 \Msun$ and $E_{51} \sim 10$).  

Computing nebular spectra for these models, we obtain synthetic spectra
resulting from various energies and ejecta masses (and thus progenitor
masses).  Summarizing, a spectrum of a particular model is specified by three
parameters: the asphericity BP, the velocity scale $f_V$ (which gives the energy in
the ejecta), and the orientation to the observer $\theta$ (defined as the angle
between the polar and the observer direction).

\subsection{Overall Spectra and Model Construction}

We compare our synthetic spectra for the two dimensional, aspherical 
supernovae with the spectra of SN 1998bw (Patat et al. 2001) obtained at 125,
200, 337, and 376 days after B maximum.  We especially focus on the +337$^{\rm
d}$ spectrum, since the epoch is sufficiently late that the contribution of
allowed Fe transitions (which are not included in the model) on the [FeII]
blend near 5200\AA\ is likely negligible (see \S 4.5.1). 

Figure 6 shows examples of the synthetic spectra with    a viewing angle
$\theta = 30^{\rm o}$.  The synthetic spectra are very much improved for the
aspherical model, especially for the one with BP=8 and $f_V$ =1.15,  compared to the
spherical hydrodynamic explosion models  (either the original CO138E30 in Fig.
1 or models BP1 in Fig. 6).  The sharply peaked MgI] 4570\AA, [OI] 6300,
6363\AA, and  NaI 5900\AA, as well as  the broad [FeII], are naturally
explained.  Indeed, the fit by the aspherical model BP8 ($f_V$=1.15) is as good as
the one by  the one dimensional "modified" CO138E30 model.  The masses of
oxygen and $^{56}$Ni are determined so as to obtain the correct line
strengths (at +337days) in the two-dimensional models.  
This is done by reducing/increasing the density uniformly 
throughout either the $^{56}$Ni or O rich region (see below).  
Therefore, each line profile 
is not sensitively affected by this procedure but results from the explosion
model itself. 

Figure 6 also shows how different kinetic energy and asphericity affect the
synthetic spectrum.  For example, we can definitely rule out the model BP2 with
$f_V$=1.6 for SN 1998bw, since the [OI] 6300\AA\ line is too broad and its shape is
less sharply peaked than in the observation.  We discuss the profiles of each
line in more detail in \S 3.3 and in the Appendix. 
The selection of models acceptable for SN 1998bw is discussed in \S 3.4.

Because the intensities of the [FeII] 5200\AA\ and the [OI] 6300\AA\ are well
defined in the observation,  they can be used to constrain the masses of iron
(mostly the product of $^{56}$Ni decay) and oxygen. These are model-dependent
as the combination of mass and energy gives the average density, which in turn
gives the deposition rate.  For example, we find that the original hydrodynamic
models (where $M_{\rm O} \sim 8 \Msun$  and $E_{51} \sim 10$) yield too strong
[OI] emission relative to [FeII] 5200\AA.  Therefore $M_{\rm O}$ should be
reduced to fit the line ratio correctly (e.g., $M_{\rm O} \sim 4.7 \Msun$ in
model BP1, $f_V$=1).  In the current study, we have uniformly reduced densities in
oxygen-dominated regions, and therefore $M_{\rm O}$, until we obtained the
correct intensity ratio [OI] 6300\AA/[FeII] 5200\AA\ at day +337.   The mass of
$^{56}$Ni is also constrained by the total luminosity in the observed spectrum.
Therefore, we also changed the mass of $^{56}$Ni (densities in $^{56}$Ni
dominated regions) until we obtained the correct luminosity at $+337^{\rm d}$. 
The masses  $M$($^{56}$Ni) and $M_{\rm O}$ consistent with the observed
flux at $+337^{\rm d}$ are thus derived for all the models as listed in Table 1.  
Note that a model with larger $f_V$ has lower 
density and smaller deposition rate, 
and therefore needs larger $M_{\rm O}$ and
$M$($^{56}$Ni) to fit the observed intensities. 
This is the reason the energy $E_{\rm K}$ in the larger $f_V$ models 
is larger than expected from the simple scaling $E_{\rm K} \propto {f_V}^2$, 
because the mass of the ejecta needed is also larger for larger $f_V$.

\subsection{Emission Lines}

Dependencies of line profiles of the [FeII] blend 5200\AA, the [OI] 6300, 6363\AA\ 
doublet, the MgI] 4570\AA, and the [CaII] 7291, 7324\AA\ on the three parameters, 
$f_V$, $\theta$, and BP are described in Appendix in detail. 
Here we give a very brief summary. 

\begin{itemize}
\item {\bf $f_V$:} Irrespective of $BP$ and $\theta$, larger $f_V$ yields 
broader profiles for every line. 
\item {\bf $\theta$:} The profile depends on whether the emitting element 
is explosive burning product (Fe and Ca) or hydrostatic product (O and Mg). 
For the former, the line is broader for smaller $\theta$. 
For the latter, the dependence is in the opposite sense. 
\item {\bf BP:} For explosive burning products, larger BP yields 
a broader [narrower] line if $\theta$ is small [large]. 
For hydrostatic products, the dependence is in the opposite sense. 
\end{itemize}

\subsection{SN 1998bw: Two Dimensional Approach}

In this section, we discuss which models are acceptable for SN 1998bw, 
according to each line profile (\S 3.3 and Appendix). 
On the basis of the selected models, we further seek for conditions 
that may yield better fits to
the observations in a manner similar to what was done for the 1D model 
(by introducing mixing and clumping; \S 2.2). 

Figure 7 illustrates how the model parameters are constrained  by the
observations.  A fit is judged by the following rules.  For the [FeII] 5200\AA,
it is checked if the width of the blend is consistent with the observed value. 
The region of acceptance is larger for $\theta = 0^{\rm o}$ than $\theta =
30^{\rm o}$, since the former results in a broader profile.  For the [OI]
6300\AA, if a model fits well the line wings and does not produce a
double-peaked emission line, it is regarded to be acceptable.  The fit to the
MgI] 4570\AA\ is not included, because the MgI] gives the same constraints as
the [OI] 6300\AA\ lines, but with less accuracy.

The fit to the [CaII] 7300\AA\ line is uncertain.  
In a strict sense, no present model, either spherical or aspherical, 
gives acceptable fits, since the observed line center is shifted blueward 
relative to the models. See Appendix for details. 
Here, we tentatively regard a fit as acceptable if the emission at the blue wing is
consistent with the observation.  
By doing this, models producing a 
doubled-peaked or a too broad flat-topped profile are excluded. 
There are several ways to shift the line
center or to depress the emission in the red, as we will discuss in \S 4. 
The region of acceptance for [CaII] 7300\AA\ moves to 
more aspherical and more energetic models (yielding a broader Ca line) for 
larger viewing angle $\theta$ (yielding a narrower Ca line), so that 
these two effects compensate one another. 

Figure 7 shows that both the [FeII] 5200\AA\ and the [OI] 6300\AA\ are
explained by highly aspherical (${\rm BP} \gsim 8$), energetic models ($f_V \gsim
1$) viewed in a direction close to the pole ($\theta \lsim 30^{\rm o}$).  
Viewing angles $\theta \gg 30^{\rm o}$ do not produce an
acceptable fit to [OI] 6300\AA.  Also, the width of the [OI] line sets an upper
limit for $f_V$ (and therefore on the energy), since models with $f_V \gsim 1.6$
produce too broad [OI] emission.  
If we include [CaII] 7300\AA\ in the fit
despite the large uncertainty, $\theta \sim 30^{\rm o}$ is preferred, since
highly aspherical and energetic models produce double peaked or too broad
[CaII] emission if viewed at $\theta\sim 0^{\rm o}$ or even at $\theta \sim 15^{\rm o}$ 
as mentioned in Maeda et al. (2002) (who did not include the [CaII] in the model). 
If only the [FeII], [OI] (and MgI]) are considered in the fit, 
the fit by models with $\theta =15^{\rm o}$ is as nice as $\theta= 30^{\rm o}$. 
The [CaII] 7300\AA\ feature is complicated and therefore needs further study 
to understand its nature (\S 4). 

Next, we examine the temporal spectral evolution of one of the acceptable
models (BP8, $f_V$=1.15, and $\theta = 30^{\rm o}$). This is shown in Figure 8a. 
Additionally, Figure 8b shows model BP8 with $f_V$=1.6 and $\theta =30^{\rm o}$.
The latter model is not regarded as "acceptable" because the [OI] 6300\AA\ line
is too broad if it is normalized at the emission peak (see Appendix). The fit
to the wings of the [OI] emission, however, is not bad, therefore the fit is
marginal. 

For SN 1998bw, this is the first attempt to check the consistency of a
theoretical model by time-dependent computations. 
Figure 8 shows that both models have difficulty in reproducing the slowly fading 
[OI] 6300\AA\ and MgI] 4570\AA\ from $+125^{\rm d}$ to $+200^{\rm d}$, 
a similar problem also encountered for the one dimensional model CO138E30.  

As done in \S 2, we have attempted to obtain better fits,  by introducing
mixing and clumping.  Guided by the 1D model, we have tried to simulate these
effects.  First, mixing between the Fe-rich region and the O-rich region is
taken into account by changing a fraction of O in the O-rich region into Fe and
Ca.  Second, we have also increased the mass of C in the O-rich region since
the synthetic feature near 8500\AA\ shows a deficiency of [CI] (Fig. 8). 
Obviously hydrodynamic explosion models, either one or two dimensional,  yield
a mass fraction of carbon in the O-rich region  insufficiently small to fit the
[CI] line strength (see also Fig. 1).  Finally, we have introduced filling
factors to simulate clumping both for the Fe-rich and the O-rich regions. 

These parameters for mixing and clumping 
are set so as to obtain a fit as nice as possible. 
Figure 9 shows the spectra of the "modified" 2D models. 
The parameters for the modification are as follows.  
For the BP8, $f_V$=1.15 model, 3\%, 1\%, and 10\% of O in the O-rich region 
are changed to Fe, Ca, and C, respectively. 
Then filling factors of 0.1 and 0.2 are introduced in the Fe- and O-rich regions, 
respectively.  For the BP8, $f_V$=1.6 model, 1.5\%, 1\%, and 10\% of
O in the O-rich region are changed to Fe, Ca, and C, respectively. Then filling
factors of 0.05 and 0.1 is introduced in the Fe- and O-rich regions,
respectively. 

Comparing Figures 8 and 9 shows that the fit is improved very much by 
introducing mixing and clumping. 
The modified 2D models match the observed spectra as nicely as 
the modified 1D model (see Fig. 9).  
The main deference between the modified 1D model and the 2D model is that   
the modified 1D model is very different from the original 
CO138E30 model. The modification yielded line profiles different from 
the original model.  
The distributions of density and abundances in this model are inconsistent 
with the results of "1D" explosion hydrodynamics. 
In the 2D model, on the other hand, the distribution 
of density and abundances are based on the aspherical explosion model. 
The changes due to the modification of the abundance distribution and the 
introduction of mixing and clumping, which may naturally occur, are small. 
This does not sensitively affect the line profiles.

\section{Summary and Discussion}

\subsection{Summary}

We have revisited the nebular spectra of SN 1998bw, using one- and
two-dimensional nebular spectrum synthesis codes.  
Compared with previous works (1D models by Sollerman et al. 2000 and Mazzali et
al. 2001, and a 2D model by Maeda et al. 2002), we have extended the analysis
in the following points.  In the 1D model, we have tried to construct a model which
explains the temporal evolution of the spectra of SN 1998bw from +125$^{\rm d}$
to +376$^{\rm d}$.  The distribution of the density and elements of the model
(the modified CO138E30) is far from what is expected from 1D spherically
symmetric explosion models.  We have further computed 2D
synthetic spectra of the aspherical explosion models of Maeda et al. (2002). 
These models show that an aspherical explosion scenario naturally 
explains the late-phase spectra of SN 1998bw.  
In the 2D model, we have extended the analysis of Maeda et al. (2002), 
including more realistic computations and more detailed comparison with the 
observations.  

As for the one-dimensional model, our main results are as follows. 
(1) The original CO138E30 model for SN 1998bw (obtained by fitting the early 
phase observations up to $\sim 2$ months) is not consistent with the late-phase 
spectra. 
(2) A high-density O-rich core below $\sim 8,000$ km s$^{-1}$ is necessary 
to explain the sharply-peaked [OI] 6300\AA, 6363\AA\ doublet profile. 
(3) At the same time, a large amount of $^{56}$Ni (which decays into Fe) should 
be located at $v \gsim 8,000$ km s$^{-1}$ to reproduce the large ratio between the 
widths of [FeII] 5200\AA\ and [OI] 6300\AA. 
The above three features are not accounted for by any spherically symmetric 
evolution-collapse-explosion calculations. 
(4) Heavy elements such as Fe and Ca are likely mixed into the O-rich 
region, and vice versa. Namely, O, Fe, and Ca do not occupy completely separated layers.   
(5) Clumping is important both in O-rich and Fe-rich regions. 

In the context of the two-dimensional models, our results can be 
summarized as follows. 
(1) The [FeII] 5200\AA\ and the [OI] 6300\AA\ lines are naturally 
reproduced by (some) aspherical models, confirming the suggestion of 
Maeda et al. (2002). 
(2) The profile of the MgI] 4571\AA\ gives additional support to the 
aspherical model for SN 1998bw. The analysis of the [CaII] profile  
demonstrates, however, that the present models are still not perfect. 
(3) Examining a wide parameter range, we have shown that nebular spectra of 
SN~1998bw indeed require a hyper-energetic explosion (see also \S 4.2). 
(4) Similar to the 1D model, mixing and clumping are at least partly 
responsible for the slow spectrum evolution from $+125^{\rm d}$ to $+200^{\rm d}$. 

\subsection{The Nature of SN 1998bw} 

According to the present model, we can infer the nature of SN 1998bw. 
It was an highly aspherical explosion (BP $\sim 8$) 
viewed at $0^{\rm o} < \theta \lsim 30^{\rm o}$. 
The viewing angle is consistent with the off-axis model ($16^{\rm o} \lsim \theta 
\lsim 36^{\rm o}$) explaing properties of prompt gamma ray emission 
of GRB 980425 associated with SN 1998bw (Yamazali, Yonetoku, \& Nakamura 2003). 
The kinetic energy in the explosion models giving the best fits 
was $E_{51} \sim 8 - 12$, mass of $^{56}$Ni $M$($^{56}$Ni) $\sim 0.4 - 0.5 \Msun$), 
and mass of oxygen in the ejecta $M_{\rm O} \sim 5 - 6 \Msun$.  
The mass of oxygen would correspond to a CO core of $\sim 9 - 10 \Msun$  
and to a He core of $\sim 11 - 12 \Msun$, which is evolved from a star with $M_{\rm ms}
\sim 30 - 35 \Msun$. 
The argument assumes the central remnant is a $4 \Msun$ black hole, which could 
be larger if a larger fraction of the O-dominated layer is accreted onto the 
black hole (e.g., MacFadyen \& Woosley 1999; Maeda \& Nomoto 2003b).  

Here we would like to discuss the uniqueness of the model. A fit to each 
line is used to restrict a possible range in the parameter space (BP, 
$f_{V}$, and $\theta$). Because the line shapes of different elements depend 
on the parameters differently (see \S 3.3 and \S 3.4), the combination of 
fits to several lines is useful to narrow down the parameter space (Fig. 7). 
Once these parameters are given, then the masses of $^{56}$Ni and ejecta mass 
are derived based on the line intensities, and therefore $M$($^{56}$Ni), 
$M_{\rm ejecta}$, and $E_{\rm K}$ are determined rather uniquely. The above 
model is derived including a qualitative fit to [CaII] 7300\AA, which is not 
very certain in the present study (see \S 3.4 and \S 4.5.2). If we omit this 
line, and use only the [OI] 6300\AA, MgI] 4570\AA, and [FeII] 5200\AA, then 
the degeneracy is not completely resolved (Fig. 7). For example, assuming 
$\theta = 30^{\rm o}$ the models with BP $\gsim 8$ and $1 \lsim f_{V} \lsim 1.45$ 
($6 \lsim E_{51} \lsim 16$) are acceptable. For smaller $\theta$, the range is 
even larger, while values $\theta \sim 0^{\rm o}$ are disfavored because they 
do not even give a qualitatively acceptable fit to [CaII] 7300\AA. The models 
can be further constrained by fitting early-phase observations, i.e., light 
curve and spectra (K. Maeda, in preparation). 

We regard the above mass of oxygen and especially the kinetic energy 
in the two-dimensional 
model as lower limits of those in the real ejecta for the following reason. 
Although we have computed a spectrum based on realistic explosion models, 
it is not certain that the models really give a good representation of material 
at velocities $\gsim 10,000$ km s$^{-1}$ since they emit little 
at the late-phases. Our explosion models contain material up to 
$\sim 15,000 - 25,000$ km s$^{-1}$ for $f_V = 1 -1.6$, above which the
density drops very rapidly as a function of radius and therefore 
was not traced by the two-dimensional explosion calculations. 
However, the early-phase spectra suggest there is material 
up to $\sim 40,000$ km s$^{-1}$ which carries a
substantial fraction of the total kinetic energy (Nakamura et al. 2001; 
see also Mazzali et al. 2000). 
Although the origin of such material is not clear, its existence suggests that 
the total kinetic energy should be larger than that of the present 
two-dimensional model. 

In any case, the lower limit of the kinetic energy $E_{51} \gsim 10$ is much
larger than the typical value $E_{51} \sim 1$. Less energetic models do not fit
the observed nebular spectra of SN 1998bw, especially at [FeII] 5200\AA. 

Also, our models suggest that the ejecta is likely very clumpy 
and that nucleosynthetically different layers are mixed. This is necessary to  
explain the spectral evolution, especially 
the slow fading of [OI] 6300\AA\ and MgI] 4570\AA\ 
between $+125^{\rm d}$ and $+200^{\rm d}$. 
The amount of mixing is small and may well be explained in the context of 
Rayliegh-Taylor or shear instabilities 
between the O-rich and the 
Fe-rich region (for Fe and Ca) and between the O-rich region and the C layer
above it (for C).  Such mixing processes are believed to take place in
core-collapse supernovae (e.g., the earlier than expected detection of X-rays 
and $\gamma$-rays from SN 1987A:  Dotani et al. 1987; Sunyaev et al. 1987; Matz
et al. 1988: see e.g., Kifonidis et al. 2000 for recent simulations). 
Also, aspherical explosions are suggested to boost the efficiency of 
the mixing (Nagataki, Shimizu, \& Sato 1998).  While we have introduced a
rather large C abundance (10\%) in the O-rich region, both C and O are
hydrostatic burning products, and therefore additional mixing could also
operate in hydrostatic evolution stages. 
This may be naturally explained by rotationally induced mixing 
(e.g., Iwamoto et al. 2005), 
since the large asphericity at the explosion implies that the progenitor 
star was a very rapid rotator.

\subsection{Implications for the Light Curve}

We derived a mass of $^{56}$Ni, $M$($^{56}$Ni) $\sim 0.4 \Msun$ in the two 
dimensional model.
This is consistent with the results of  previous work (e.g., Nakamura et al.
2001).  The distribution of $^{56}$Ni in our model is characterized by a large
amount of $^{56}$Ni in the high velocity outer region, and a small fraction in
the inner high density region.  The distribution will affect the shape of the
light curve.  Figure 10 shows the evolution of the luminosity deposited by
$\gamma$-rays and positrons, i.e., optical light curve applicable after 
$\sim 50 - 100$ days. 
At least qualitatively, the 2D models are favored over the original 1D CO138E30 
model, thanks to the presence of $^{56}$Ni at high velocities and 
the high density core at low velocities (Maeda et al. 2003a). 
Computation of 2D light curves from the explosion to the late-phases 
on the basis of the two dimensional models will be another check of the 
validity of the models (see also H\"oflich, Wheeler, \& Wang 1999). 

\subsection{Notes for Future Observations} 

Our suggestion that hypernovae are aspherical explosions may 
further be able to be confirmed by future observations.   
First, we suggest to take late-phase near-infrared (NIR) spectra of 
supernovae with SN 1998bw-like early-phase spectra, as has been done for SNe Ia 
(e.g., H\"oflich et al. 2004). 
In the NIR, a [FeII] line is more isolated than in the optical band, 
making the effect of the orientation easier to see. 

It is also interesting to examine the [OI] 6300\AA\ profile in other hypernovae
and SNe Ib/c to investigate the effect of the viewing angle.  Recently,
Kawabata et al. (2004) reported the detection of the double-peaked [OI]
6300\AA\ in a spectrum of SN 2003jd at $\sim 1$ year after the discovery. 
Mazzali et al. (2005) interpreted it as a SN 1998bw-like event, as suggested
from  early-phase spectroscopy, but viewed from the equatorial direction. 
Although the sample is small in number for hypernovae, 
late-phase spectra such as those published
for SNe Ib/c by Matheson et al. (2001) would allow statistical studies to
constrain the explosion energy and the asphericity of SNe Ib/c.

\subsection{Remaining Problems} 

\subsubsection{The Spectrum at $+125^{\rm d}$}

One of the major advances in the present study is the time-dependent
computations of the nebular spectra of SN 1998bw.  The model derived at the
spectrum at +337$^{\rm d}$ gives nice fits to the sequence of the spectra after
+200$^{\rm d}$.  
However, at the first epoch +125$^{\rm d}$, some deviation from
the observations still exists especially in the luminosity of the [OI] 6300,
6363\AA\ doublet, even though we have tried to fix this 
by introducing mixing and clumping. 
Others are the strong [FeII] and [FeIII] between 4000\AA\ and 5000\AA\, and the strong
MgI] 4570\AA\ and [OI] 5577\AA\ in the model.  They may simply be due to inappropriate
treatment of the first epoch +125$^{\rm d}$ in our computation since at this
epoch the ejecta density is still high and the nebular representation may not 
totally be a good approximation. 
Temporal evolution of the intensity of the [OI] 6300\AA\ changed around 
$+125^{\rm d}$ (Patat et al. 2001), implying that at this epoch the ejecta are 
still in the transition from the photospheric phase to the nebular phase. 
For example, the observationally strong OI 7800\AA\ implies that 
allowed transitions may be dominant in some wavelength ranges. 

Because of the possible presence of emission lines of some strong allowed Fe II 
transitions, the identification of the 5200\AA\ feature as the blend of forbidden 
lines only may be uncertain. Axelrod (1980) computed optical depths of the strongest 
allowed Fe II lines near 5200\AA\ for a model SN Ia nebula at days 87 and 264 to 
be of the order of 10 and 0.1, respectively. Given the larger ratio 
$E_{\rm K}/M$($^{56}$Ni) in our models for SN 1998bw than for SNe Ia, the optical 
depths will be even smaller. Rough estimates suggest that a few lines may have 
optical depth of order unity at the first epoch ($+125^{\rm d}$). Therefore, at 
this epoch the contribution of the allowed Fe II lines and related radiation 
transfer effects, e.g., line scattering, may affect (but probably not dominate) 
the shape of the 5200\AA\ feature. At the same time, the same estimate shows that 
the optical depths of the allowed Fe lines will be very small (of the order of 0.1 
or less) at $+200^{\rm d}$ and thereafter, justifying the assumption that the 
contribution of the allowed lines and radiation transfer effects are negligible. 
See also Maeda et al. (2002) for a detailed discussion of the identification of 
the 5200\AA\ feature as the forbidden lines. 

As for the MgI] 4570\AA, 
the line emissivity is quite sensitive to the treatment of the photoionization
radiation field because of its low ionization threshold (Houck \& Fransson
1996; Kozma \& Fransson 1998a), while the ionization by UV photons is not
included in our present spectrum synthesis calculations (See also, e.g.,
Sollerman et al. 2004 for the effect of photoionization).  This could partly be
a reason of our failure in fitting the temporal evolution of the MgI] 4570\AA\
luminosity, since the effect is expected to be stronger at earlier epochs. 
We plan to extend our code to allow better treatment at 
such relatively early epochs.

\subsubsection{Peculiar [CaII] Emission}

The final question we should answer in the future concerns the origin of the
peculiar [CaII] 7300\AA\ emission.  As mentioned in Appendix, the model spectra
are too red.  It is interesting that another
hypernova, SN 2002ap, shows the [CaII] 7300\AA\ line exactly at the correct
position (Kawabata et al. 2002; Leonard et al. 2002; Wang et al. 2003; Foley et
al. 2003). The blueshift of the line in SN 1998bw seems to be unique.  

The feature is a complex blend and therefore it is difficult to 
clarify the reason of the failure in the model. 
With this caveat taken in mind, we speculate possibilities 
which may explain the discrepancy. 
The problem is possibly related to the distribution of Ca. 
There are at least two possibilities to reconcile the problem. 
First, the distribution of Ca may be similar to O, rather than Fe. 
It will then be challenging to the theory of explosive nucleosynthesis. 
Another possible interpretation would be that 
SN1998bw may have asymmetry even between the two hemispheres.  If we look at
the event from the hemisphere with the larger kinetic energy,  the center of
the Ca distribution would be blue shifted.  This should however also apply to
[FeII] 5200\AA, which does not seem to show the shift.  The significantly
blended  nature of [FeII] 5200\AA\ may wash that signal away.  In this context,
SN2002ap would be an explosion in which  the degree of this asymmetry is small
or the viewing orientation is relatively large.  This is an interesting
possibility, and we will pursue this issue in the future. 
The above two possibilities could be resolved by examining 
a number of hypernova late-phase spectra. In our interpretation,  
the sample contains hypernovae with various viewing angles, then 
the two different Ca distributions -- oxygen-like or iron-like -- should 
give different statistics.

\acknowledgements

This work has been supported in part by the grant-in-Aid for Scientific
Research and the 21st Century COE program of the Ministry of
Education, Culture, Sports, Science and Technology in Japan. 
KM is Research Fellow of the Japan Society for the Promotion of Science  
(JSPS). 

\appendix

\section{Emission Line Profiles}

\subsection{[FeII] 5200\AA}

Figures 11, 12, and 13 show how the shape of the [FeII] blend at 5200\AA\ depends
on the degree of asphericity and the observer direction.  The dependence on the
kinetic energy can also be seen by comparing Figures 11 ($f_V$=1.6), 12 ($f_V$=1.15), and
13 ($f_V$=0.7). 

The dependencies can be summarized as follows. 
(1) For a given asphericity, a smaller viewing angle $\theta$ leads to a 
broader feature (with the obvious exception of the spherically symmetric 
models). 
(2) Larger asphericity leads to a broader [narrower] 
feature for small [large] $\theta$. 
(3) Larger kinetic energy leads to a broader feature irrespective 
of the degree of asphericity and orientation. 

The above dependencies can be understood from the distribution of $^{56}$Ni
(Fe) in Figure 5.  The distribution is elongated in the polar direction in the
aspherical models. Larger asphericity leads to higher velocities in the polar
direction and smaller ones in the equatorial direction.  If the energy is
larger, the velocity in all directions becomes larger.  These characters of the
$^{56}$Ni distribution explain the dependence of the [FeII] feature on various
parameters.  

The observed broad 5200\AA\ emission can be explained if the asphericity is
large, the energy is large, and the viewing angle $\theta$ is small ($\theta
\lsim 30^{\rm o}$).  For example, Figure 11 shows that for these very energetic
models ($f_V$=1.6), the synthetic [FeII] 5200\AA\ is as broad as the observed one
for BP $\gsim 2$ given $\theta \lsim 30^{\rm o}$.  For less energetic models,
the criterion is tighter:  For $f_V$=1.15, BP $\gsim 8$ is necessary to produce the
broad feature.  For $f_V$=0.7, the synthetic [FeII] 5200\AA\ is never as broad as 
the observed one, even if the asphericity BP is very large.  
Note also that if both the
energy and the asphericity are extremely large (e.g., $f_V$=1.6 and BP=16), the
synthetic feature is too broad.

\subsection{[OI] 6300, 6363\AA}

The dependence of the [OI] 6300, 6363\AA\ doublet profile on various parameters
is shown in Figures 14 -- 16.  The observed sharply-peaked profile with
extended wings again favors highly aspherical models (BP16 or BP8) viewed from
a direction close to the pole ($\theta \lsim 30^{\rm o}$).  In addition, the
narrow [OI] 6300\AA\ sets constraint on the model in a different way from the
way the [FeII] 5200\AA\ line does.  For the less energetic models ($f_V$=0.7), all
the aspherical models (BP = 2 -- 16) give nicely peaked profiles 
if viewed near the pole.  For $f_V$=1.15, larger BP (BP $\gsim 4$) is
required.  Furthermore, for $f_V$=1.6, the synthetic [OI] 6300\AA\ is broader than
the observations, even for small viewing angles. 

The sharply-peaked [OI] profile is produced by the edge-on view of a disk-like
oxygen distribution. The model predicts a double-peaked profile if viewed near
the equator as shown in Figures 14 -- 16 for large $\theta$.  A spherical
explosion model like CO138E30 produces a flat-topped profile because of the
central hole in oxygen  distribution (at $\sim 10,000$ km s$^{-1}$ for the
kinetic energy $E_{51} = 30$;  Figs. 4a and 5), which does not fit well the
observations, irrespective of the total kinetic energy. This is especially
evident in the larger kinetic energy model ($f_V$ $\gsim 1$) which produces very
broad [OI] 6300\AA\ emission.

\subsection{MgI] 4570\AA}

The observed profile of MgI] 4570\AA\ is sharply peaked, 
similar to [OI] 6300\AA\ (Patat et al. 2001). 
The computation of nucleosynthesis in a supernova explosion gives a
distribution of Mg similar to that of O.  Therefore, the line profile 
shows characteristics similar to those of O.  
Figures 17 -- 19 show the synthetic line profiles of MgI] 4570\AA\ for various
models and viewing angles.  The fit to this line is less certain than that of
[OI] 6300\AA\, because of significant contributions of the [FeII] and [FeIII]
lines.  To examine the characteristics of pure MgI] 4570\AA, the contribution
of this line is also shown in Figures 17 -- 19.  The shape of the line favors
highly-aspherical models viewed at small $\theta$, supporting the results 
derived from the [OI] 6300\AA\ profile.

\subsection{[CaII] 7291, 7324\AA}

Figures 20 -- 22 show synthetic spectra of the present models around 7300\AA\ 
compared to the observed one at 337 days after the B maximum.  The feature is a
blend of [CaII] 7291\AA, 7324\AA, [FeII] 7155\AA, 7172\AA, 7388\AA, and
7452\AA, [CoII] 7541\AA, and [OII] 7322\AA. 
In a typical nebular condition, the [CaII] lines (whose contributions are
individually shown in Figures 20 -- 22) probably dominate at the line center,
while [FeII] (and [CoII]) fills up the wings.  The [OII] 7322\AA\ line is weak
in the model, because most oxygen is neutral, as suggested by the strong [OI]
6300\AA\ line. 

As shown in the figures, the models are redder than the observations.  
As mentioned above, the feature is a complex blend making the reason of the failure 
difficult to identify. 
For possible origins of the blueshift, see \S 4.  

The [CaII] 7292\AA\ and 7324\AA\ in the models BP16 with small viewing angle 
($\theta \sim 0^{\rm o}$) show double peaked profiles because the distribution
of Ca, as a product of explosive nucleosynthesis, follows closely that of Fe
rather than that of O, which is a product o f hydrostatic burning and a fuel 
at the explosion (Fig. 5).  
A higher degree of asphericity (i.e., larger BP) leads to more aspherically
distributed Ca.  Therefore, the double-peaked profile of Ca for small viewing
angle ($\theta \sim 0^{\rm o}$) can be interpreted as two-blobs of Ca moving in
opposite directions observed on the symmetry axis.  We do not see such a
profile in SN 1998bw, suggesting a slightly off-axis viewing angle.   

Because of the significant blend of the feature at 7300\AA, constraining the
energy in the ejecta by fitting this feature is very uncertain.  As long as 
only line width is concerned, less energetic models are preferred if $\theta$
is smaller (e.g., $\theta = 0^{\rm o}$), and more energetic ones if $\theta$ is
larger (e.g., $\theta = 30^{\rm o}$), since narrowing [broadening] the line by
smaller [larger] energy  must be compensated by  broadening [narrowing] the
line by smaller [larger] $\theta$.

\onecolumn

\clearpage
\begin{deluxetable}{cccccc}
 \tabletypesize{\scriptsize}
 \tablecaption{Models
\label{tab:model}}
 \tablewidth{0pt}
 \tablehead{
   \colhead{Model\tablenotemark{a}}
 & \colhead{Asphericity (BP)}
 & \colhead{$f_V$\tablenotemark{b}}
 & \colhead{${\KE}/10^{51}$ ergs}
 & \colhead{$M$($^{56}$Ni)$/\Msun$}
 & \colhead{$M_{\rm O}/\Msun$} 
 }
\startdata
BP16   & 16  & 1.9  & 33.1& 0.79 & 7.6\\
       & 16  & 1.75 & 26.5& 0.67 & 7.0\\
       & 16  & 1.6  & 20.7& 0.63 & 6.4\\
       & 16  & 1.45 & 15.9& 0.60 & 5.8\\
       & 16  & 1.3  & 11.5& 0.52 & 5.0\\
       & 16  & 1.15 & 8.2 & 0.45 & 4.4\\
       & 16  & 1.0  & 5.9 & 0.34 & 4.1\\
       & 16  & 0.85 & 3.8 & 0.27 & 3.3\\
       & 16  & 0.7  & 2.3 & 0.20 & 2.8\\
BP8    &  8  & 1.9  & 34.5& 0.75 & 8.7\\
       &  8  & 1.75 & 27.5& 0.68 & 8.1\\
       &  8  & 1.6  & 21.3& 0.63 & 7.3\\
       &  8  & 1.45 & 15.3& 0.55 & 6.1\\
       &  8  & 1.3  & 11.5& 0.48 & 5.6\\
       &  8  & 1.15 & 8.4 & 0.37 & 5.1\\
       &  8  & 1.0  & 5.9 & 0.32 & 4.6\\
       &  8  & 0.85 & 4.0 & 0.25 & 4.1\\
       &  8  & 0.7  & 2.0 & 0.22 & 2.6\\ 
BP4    &  4  & 1.9  & 32.5& 0.63 & 7.9\\
       &  4  & 1.75 & 25.9& 0.58 & 7.3\\
       &  4  & 1.6  & 20.2& 0.53 & 6.7\\
       &  4  & 1.45 & 15.3& 0.45 & 6.0\\
       &  4  & 1.3  & 11.2& 0.42 & 5.3\\
       &  4  & 1.15 & 7.9 & 0.36 & 4.5\\ 
       &  4  & 1.0  & 5.7 & 0.30 & 4.2\\ 
       &  4  & 0.85 & 3.7 & 0.25 & 3.5\\ 
       &  4  & 0.7  & 2.1 & 0.19 & 2.6\\
BP2    &  2  & 1.9  & 36.9& 0.55 & 9.7\\
       &  2  & 1.75 & 28.9& 0.51 & 8.7\\
       &  2  & 1.6  & 22.1& 0.47 & 7.8\\
       &  2  & 1.45 & 16.3& 0.43 & 6.7\\
       &  2  & 1.3  & 12.2& 0.36 & 6.1\\
       &  2  & 1.15 & 8.8 & 0.33 & 5.5\\
       &  2  & 1.0  & 6.2 & 0.27 & 4.9\\
       &  2  & 0.85 & 4.1 & 0.24 & 4.3\\
       &  2  & 0.7  & 2.9 & 0.22 & 3.4\\
BP1    &  1  & 1.9  & 37.5& 0.54 & 9.5\\
       &  1  & 1.75 & 29.7& 0.49 & 8.6\\
       &  1  & 1.6  & 23.1& 0.47 & 7.7\\
       &  1  & 1.45 & 16.7& 0.43 & 6.4\\
       &  1  & 1.3  & 12.7& 0.41 & 5.8\\
       &  1  & 1.15 & 9.5 & 0.39 & 5.4\\
       &  1  & 1.0  & 6.6 & 0.33 & 4.7\\
       &  1  & 0.85 & 4.1 & 0.30 & 3.5\\
       &  1  & 0.7  & 2.5 & 0.22 & 2.9\\
\enddata
\tablenotetext{a}{The models corresponds to the ones in Maeda et al (2002) as 
follows: BP16 (A), BP8 (C), BP2 (E), BP1 (F: a spherical model). 
The models BP4 are newly calculated in the way same with Maeda et al. (2002).}
\tablenotetext{b}{Velocities at any points in the ejecta are multiplied 
by a factor of $f_V$. $f_V = 1$ for the original models in Maeda et al. (2002).}
\end{deluxetable}

\clearpage
\begin{figure}
\begin{center}
		\epsscale{0.6}
		\plotone{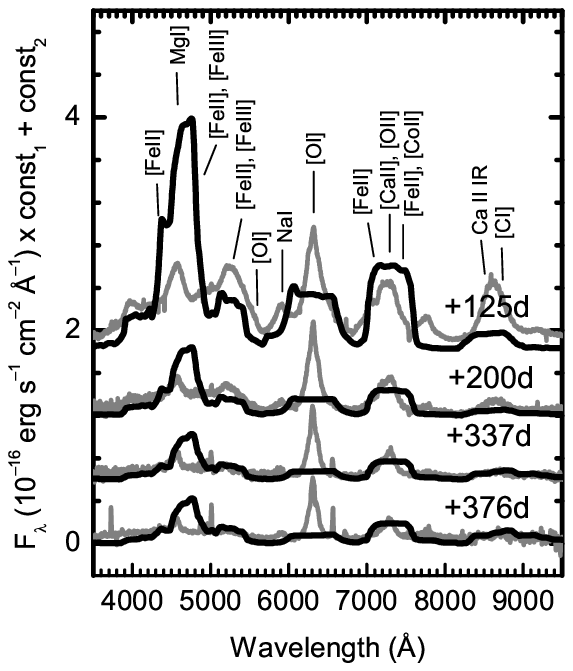}
\end{center}
\caption[]
{Synthetic spectra (black) for the original CO138E30 model as 
compared with the spectra of SN 1998bw (gray). 
For presentation, both the synthetic and observed spectra 
are multiplied by an arbitrary constant (const$_{1}$) and 
then vertically shifted upward by another factor (const$_{2}$). 
The amounts (const$_{1}$, const$_{2}$ in units of $10^{-16}$ 
erg s$^{-1}$ cm$^{-2}$ \AA$^{-1}$) are as follows: 
+125$^{\rm d}$ (0.1, 1.8), +200$^{\rm d}$ (0.1, 1.2), 
+337$^{\rm d}$ (0.5, 0.6), and +376$^{\rm d}$ (1.0, 0.0). 
The mass of $^{56}$Ni is $0.8 \Msun$. 
\label{fig1}}
\end{figure}

\clearpage
\begin{figure}
\begin{center}
		\epsscale{0.6}
		\plotone{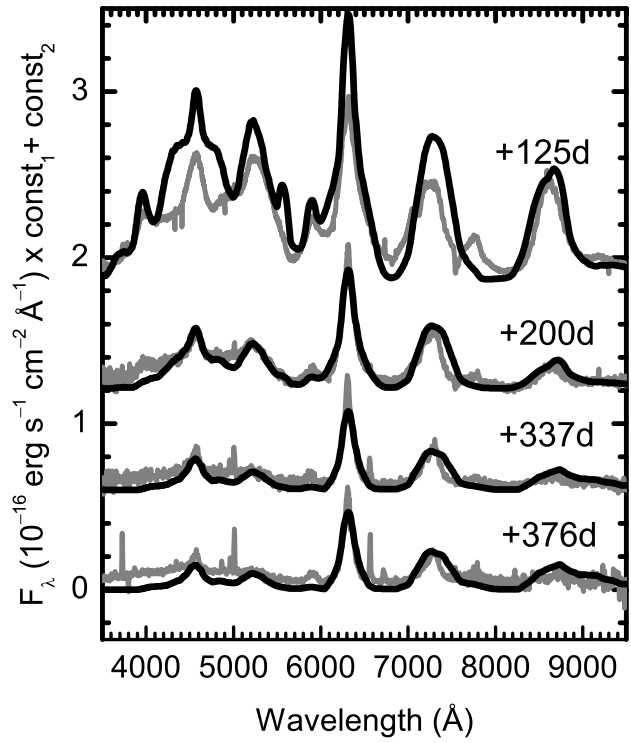}
\end{center}
\caption[]
{Synthetic spectra for the modified CO138E30 model.  
See the caption of Fig. 1. The mass of $^{56}$Ni is 
$0.61 \Msun$. 
\label{fig2}}
\end{figure}

\clearpage
\begin{figure}
\begin{center}
		\epsscale{0.6}
		\plotone{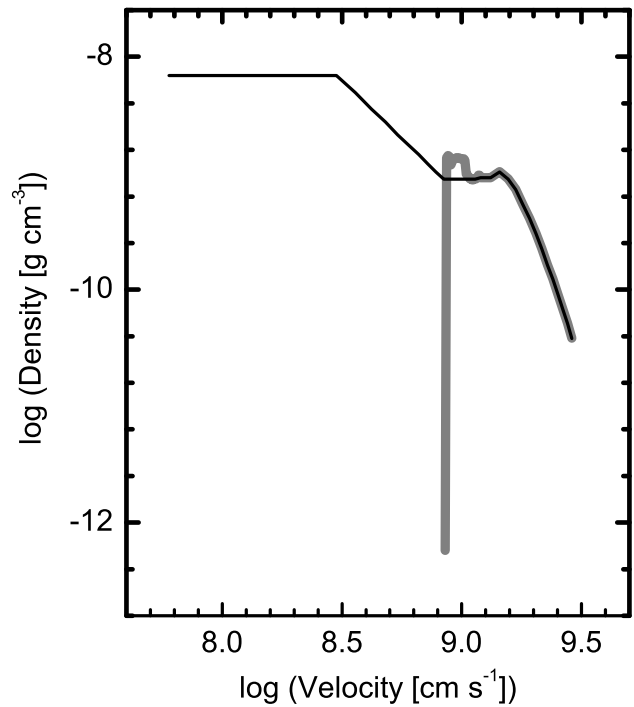}
\end{center}
\caption[]
{Density distribution at 1 day after the explosion. 
The original CO138E30 (gray) and 
the modified model (black) are shown. 
High velocity materials at 
$> 30,000$ km s$^{-1}$ are not included in the 
present study, since they contribute virtually nothing 
on late phase emission. 
\label{fig3}}
\end{figure}

\clearpage
\begin{figure}
\begin{center}
		\epsscale{0.6}
		\plotone{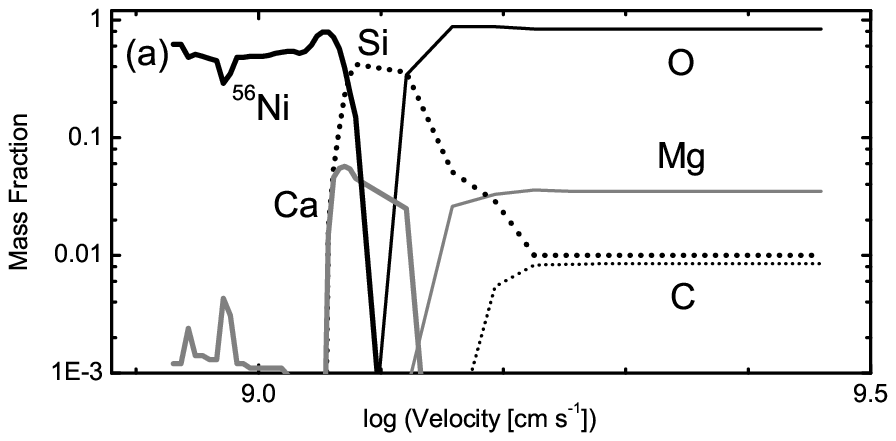}
		\epsscale{0.6}
		\plotone{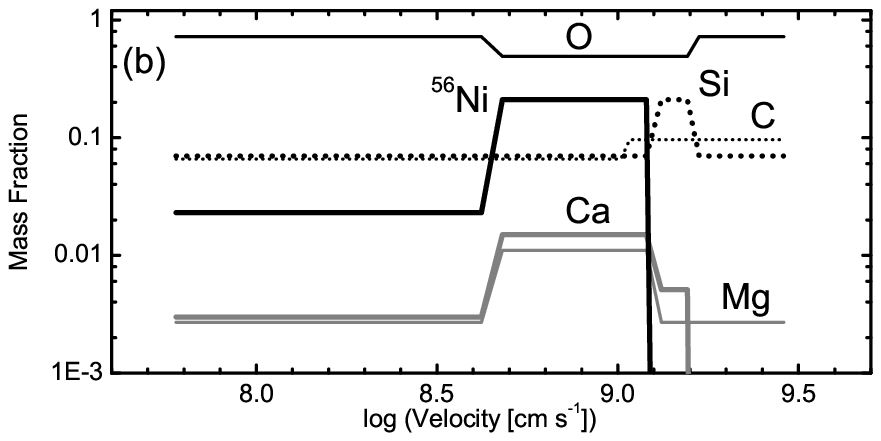}
\end{center}
\caption[]
{Mass fractions of some selected elements and $^{56}$Ni in 
the velocity space for (a) the original CO138E30 and for 
(b) the modified model. 
\label{fig4}}
\end{figure}

\clearpage
\begin{figure}
\begin{center}
		\epsscale{.7}
		\plotone{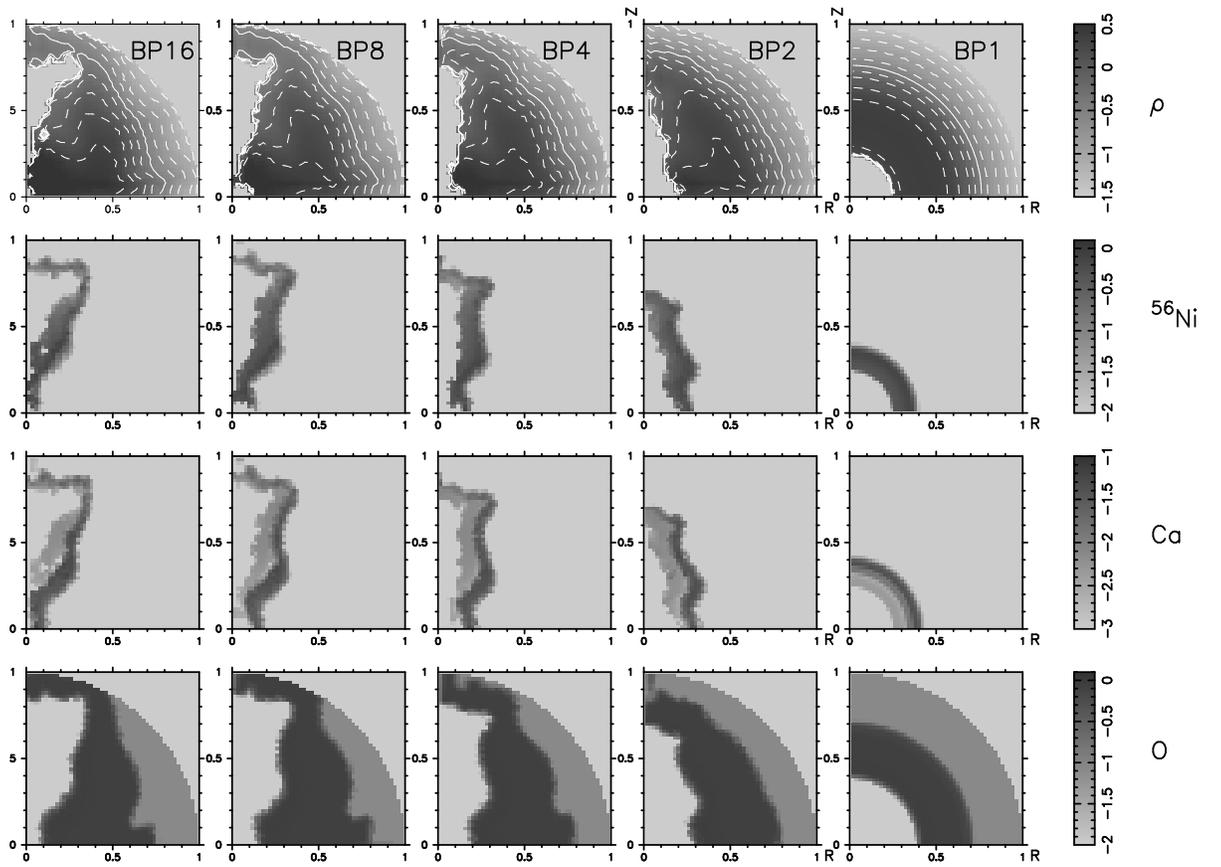}
\end{center}
\caption[]
{Structure of the ejecta for Models BP16, BP8, BP4, BP2, 
and BP1 (from left to right) in the ejecta velocity space at 
a reference time (at 100 seconds). The parameter 
$f_V$ = 1. The axes are $r$ and $z$ components of velocities 
scaled at $15,000$ km s$^{-1}$.  
For each model, shown here are density, 
mass fraction of $^{56}$Ni, Ca, and O (from top to bottom).
\label{fig5}}
\end{figure}

\clearpage
\begin{figure}
\begin{center}
		\epsscale{.45}
		\plotone{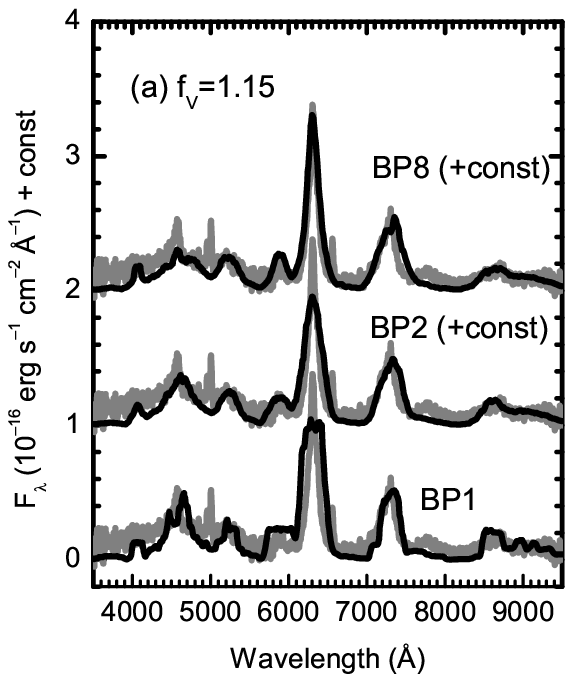}
		\epsscale{.45}
		\plotone{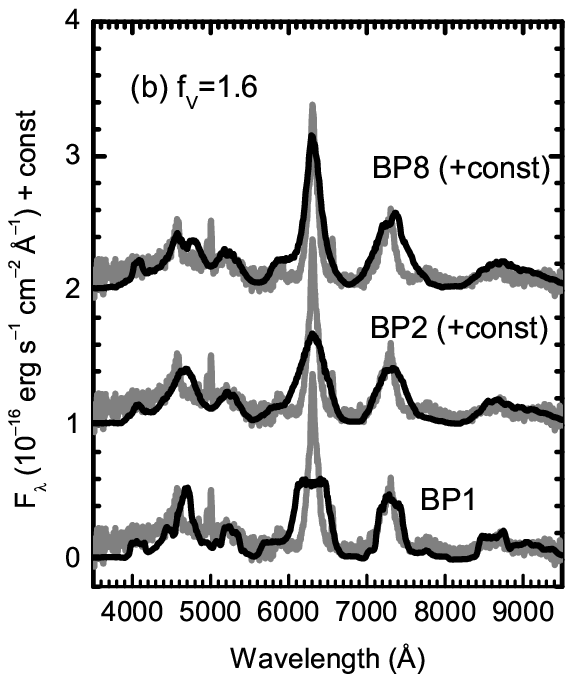}
\end{center}
\caption[] 
{Examples of synthetic spectra (black lines) as compared with 
the spectrum of SN 1998bw at 337 days after the maximum light
(gray lines). 
Shown here are Models BP8, BP2, and BP1 with (a) $f_V$=1.15 and 
$\theta = 30^{\rm o}$ and with (b) $f_V$=1.6 and $\theta = 30^{\rm o}$. 
The spectra of BP8 and BP2 are shifted vertically for presentation 
(+2 and +1 $\times 10^{-16}$ erg s$^{-1}$ cm$^{-2}$ \AA$^{-1}$) for BP8 and 
BP2, respectively). 
\label{fig6}}
\end{figure}

\clearpage
\begin{figure}
\begin{center}
		\epsscale{.5}
		\plotone{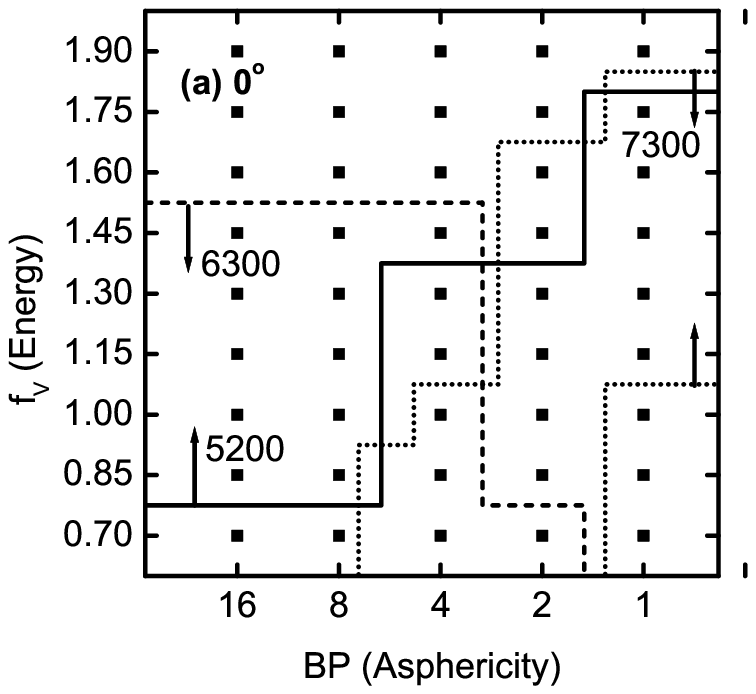}
		\epsscale{.5}
		\plotone{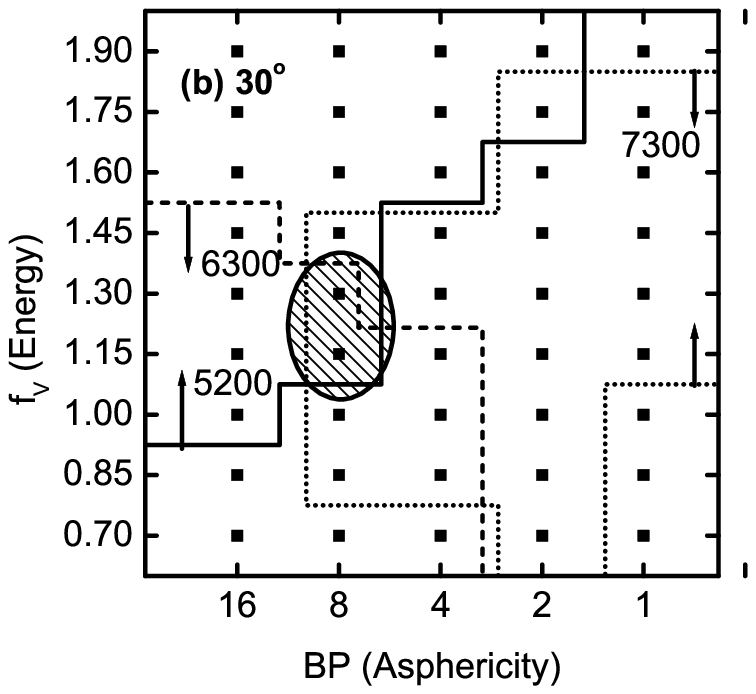}
\end{center}
\caption[] 
{Acceptable models for 1998bw. 
(a) $\theta = 0^{\rm o}$ and (b) $\theta = 30^{\rm o}$.  
The figure is obtained by comparing model spectra with 
the observed one of SN 1998bw at 337 days after the B maximum. 
The regions surrounded by solid, dashed, and dotted lines contain 
models which give an acceptable fit to the line profiles 
around 5200\AA\ ([FeII]), 6300\AA\ ([OI]), and 7300\AA\ ([CaII]), 
respectively. See \S 3.4 for details. 
\label{fig19}}
\end{figure}

\clearpage
\begin{figure}
\begin{center}
		\epsscale{.45}
		\plotone{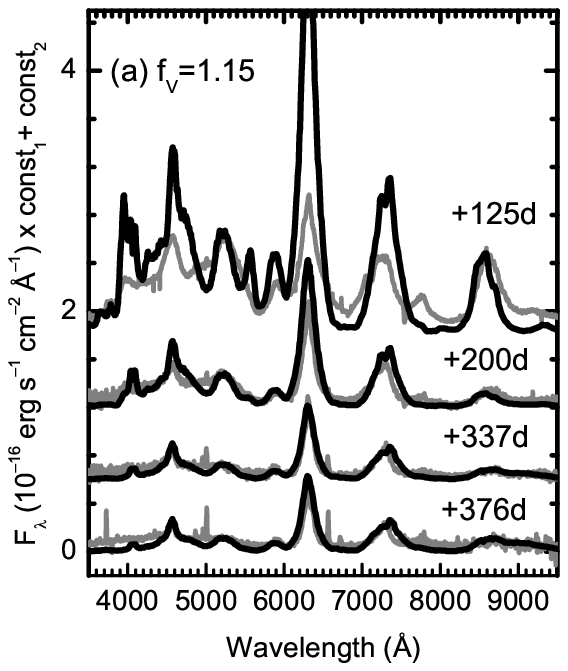}
		\epsscale{.45}
		\plotone{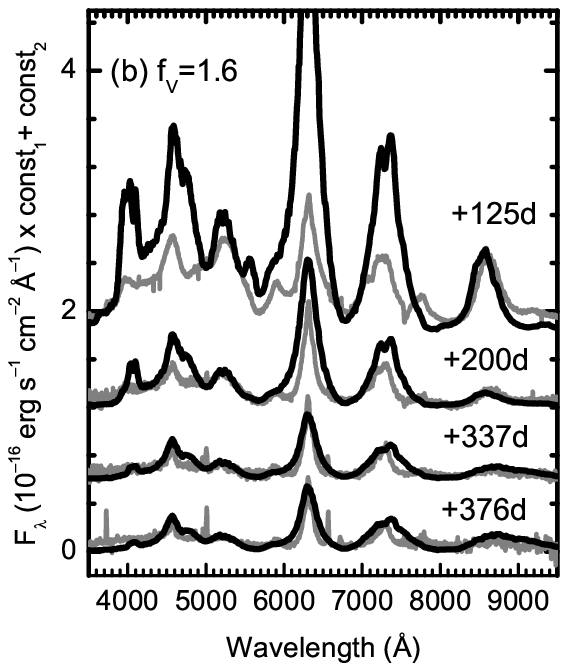}
\end{center}
\caption[] 
{Temporal evolution of the synthetic spectra for 2D models as 
compared with the spectra of SN 1998bw. 
Shown here are Model BP8 with $f_V$=1.15 (a) and $f_V$=1.6 (b). 
The orientation is $\theta = 30^{\rm o}$. 
For presentation, both the synthetic and the observed spectra 
are multiplied by an arbitrary constant (const$_{1}$) and 
then vertically shifted upward by another factor (const$_{2}$). 
The amounts (const$_{1}$, const$_{2}$ in units of 
$10^{-16}$ erg s$^{-1}$ cm$^{-2}$ \AA$^{-1}$) are as follows: 
+125$^{\rm d}$ (0.1, 1.8), +200$^{\rm d}$ (0.1, 1.2), 
+337$^{\rm d}$ (0.5, 0.6), and +376$^{\rm d}$ (1.0, 0.0). 
\label{fig20}}
\end{figure}

\clearpage
\begin{figure}
\begin{center}
		\epsscale{.45}
		\plotone{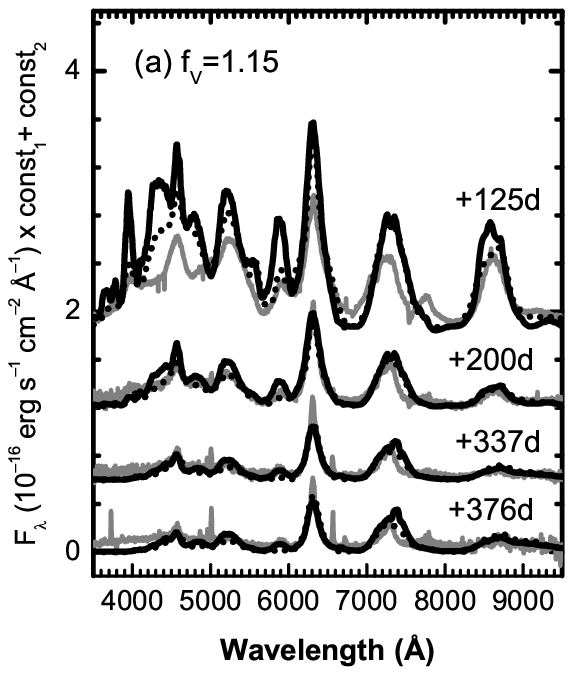}
		\epsscale{.45}
		\plotone{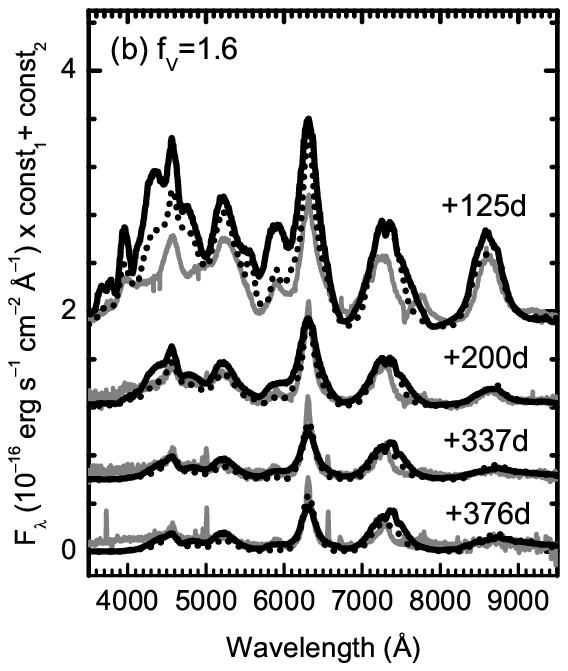}
\end{center}
\caption[] 
{Sequence of the spectra of the "modified" two-dimensional models 
(black solid lines).  
Shown here are Model BP8 with $f_V$=1.15 (a) and $f_V$=1.6 (b). 
The orientation is $\theta = 30^{\rm o}$. 
The spectra of SN 1998bw (gray lines) and the modified "one dimensional" 
CO138E30 model (black dotted line) are also shown. 
See also the caption of Figure 8. 
The mass of $^{56}$Ni in the modified models is 
$0.40 \Msun$ for $f_V$=1.15 (a) and $0.56 \Msun$ for $f_V$=1.6 (b), respectively. 
\label{fig21}}
\end{figure}

\clearpage
\begin{figure}
\begin{center}
		\epsscale{0.6}
		\plotone{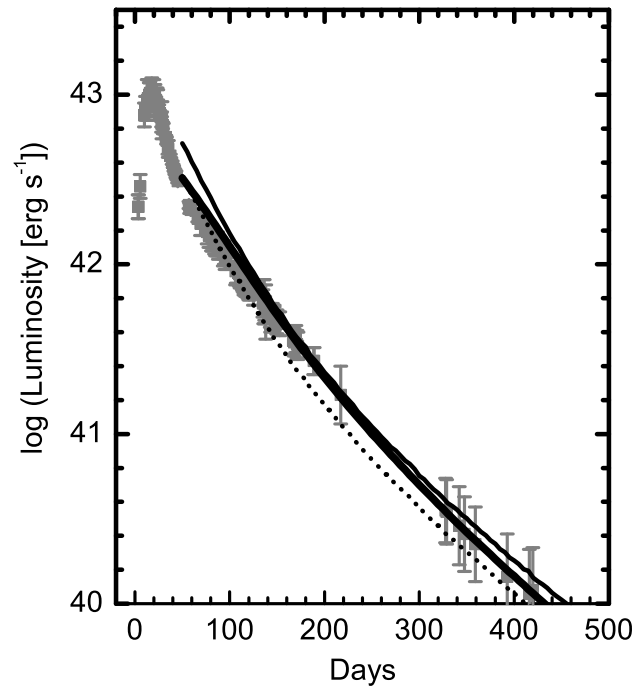}
\end{center}
\caption[] 
{The bolometric light curve of SN 1998bw (gray points; Patat et al. 2001) 
are compared with the synthetic curve from the 1D original CO138E30 
(thin lines) and that from the 2D BP8 model with $f_V$=1.15 (thick solid). 
The mass of $^{56}$Ni in the 2D model is $0.37 \Msun$ (Table 1). 
In the 1D model, the mass of $^{56}$Ni is varied to illustrate the 
difference with the 2D model. $M$($^{56}$Ni) is $0.64 \Msun$ (thin solid) 
and $0.41 \Msun$ (thin dotted) 
for the 1D model.  
\label{fig23}}
\end{figure}

\clearpage
\begin{figure}
\begin{center}
		\epsscale{1.0}
		\plotone{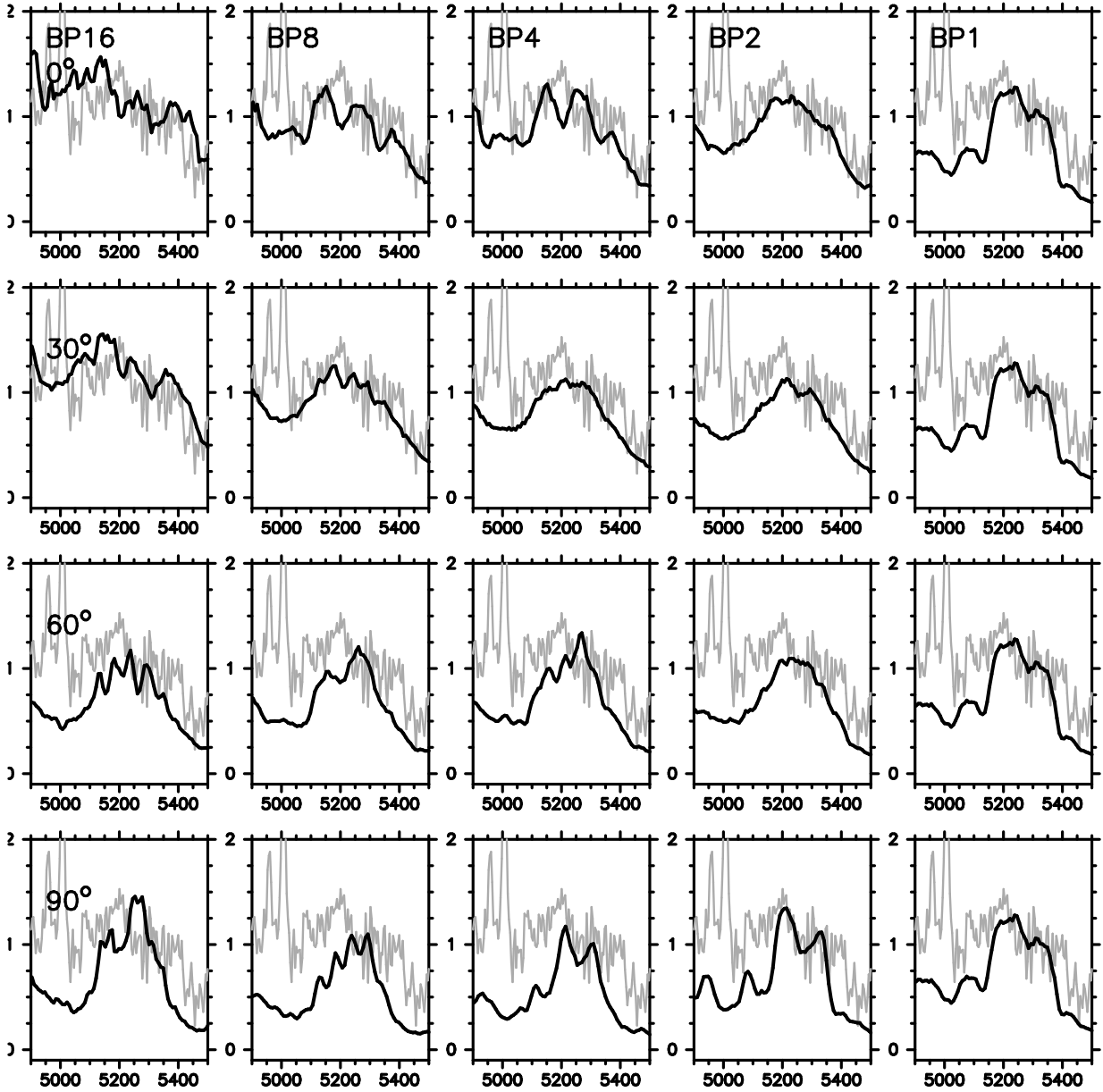}
\end{center}
\caption[] 
{The synthetic spectra of the 2D models 
around 5200\AA\ (black lines) 
as compared with the observed one of SN 1998bw at 337 days 
after the B maximum (gray lines). 
The synthetic spectra are mainly contributed by [FeII]. 
The models are those with $f_V=1.6$.
Shown here are BP16, 8, 4, 2, 1 (from left to right) 
and $\theta = 0, 30, 60, 90^{\rm o}$ (from top to bottom). 
\label{fig7}}
\end{figure}

\clearpage
\begin{figure}
\begin{center}
		\epsscale{1.0}
		\plotone{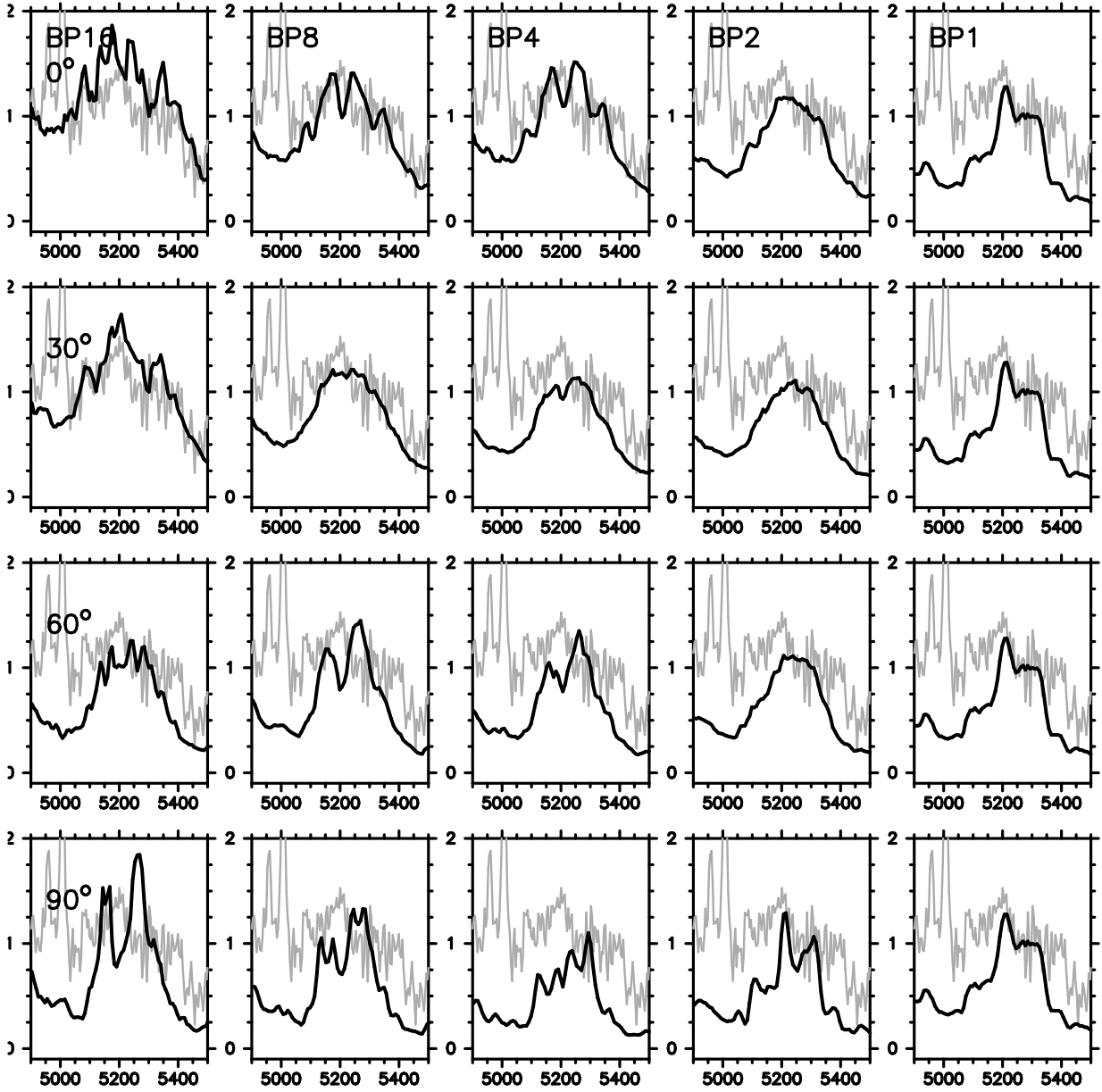}
\end{center}
\caption[] 
{The same with Fig. 11 ([FeII] 5200\AA), except 
the model parameter $f_V=1.15$. 
\label{fig8}}
\end{figure}

\clearpage
\begin{figure}
\begin{center}
		\epsscale{1.0}
		\plotone{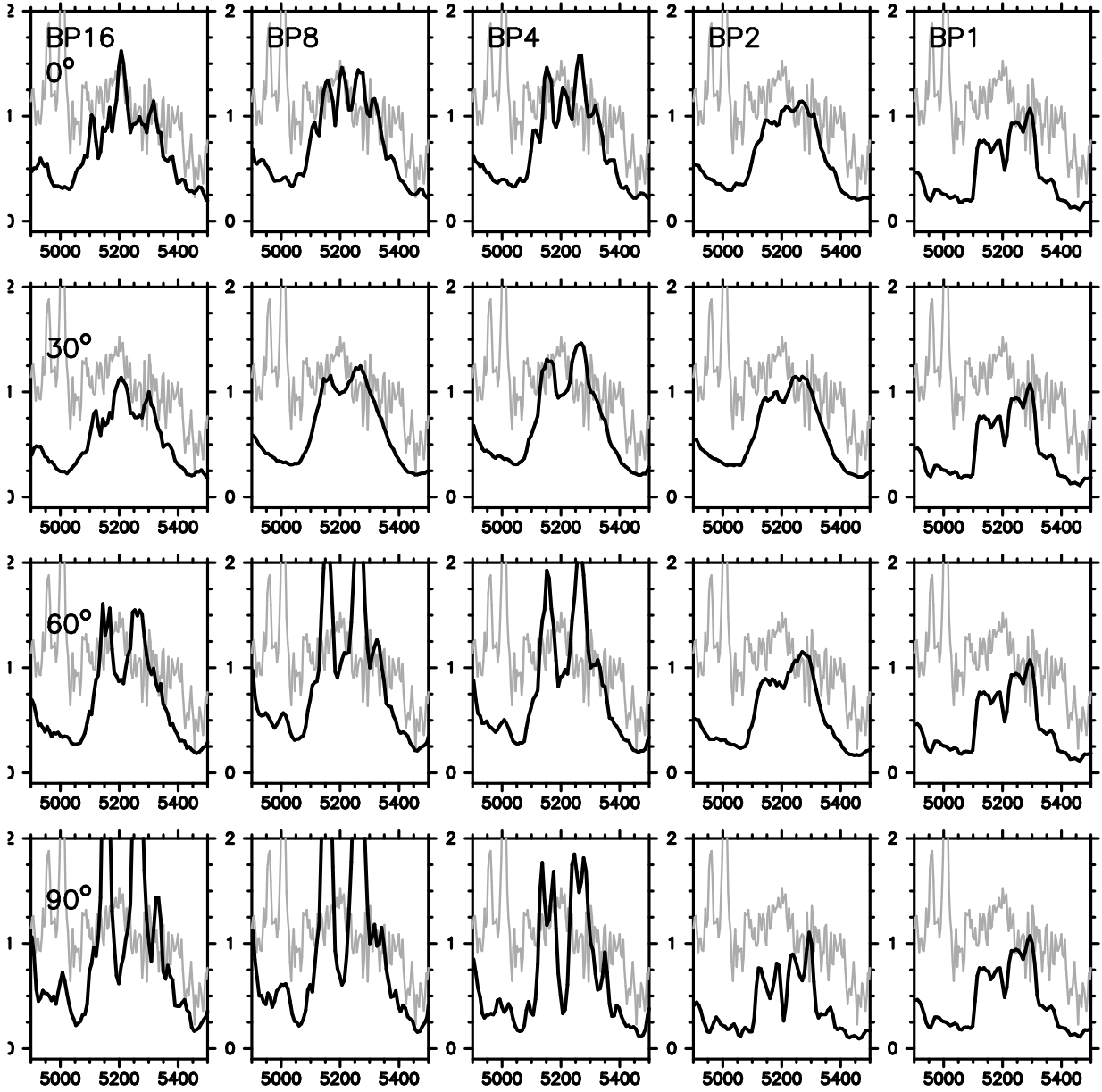}
\end{center}
\caption[] 
{The same with Fig. 11 ([FeII] 5200\AA), except 
the model parameter $f_V=0.7$. 
\label{fig9}}
\end{figure}

\clearpage
\begin{figure}
\begin{center}
		\epsscale{1.0}
		\plotone{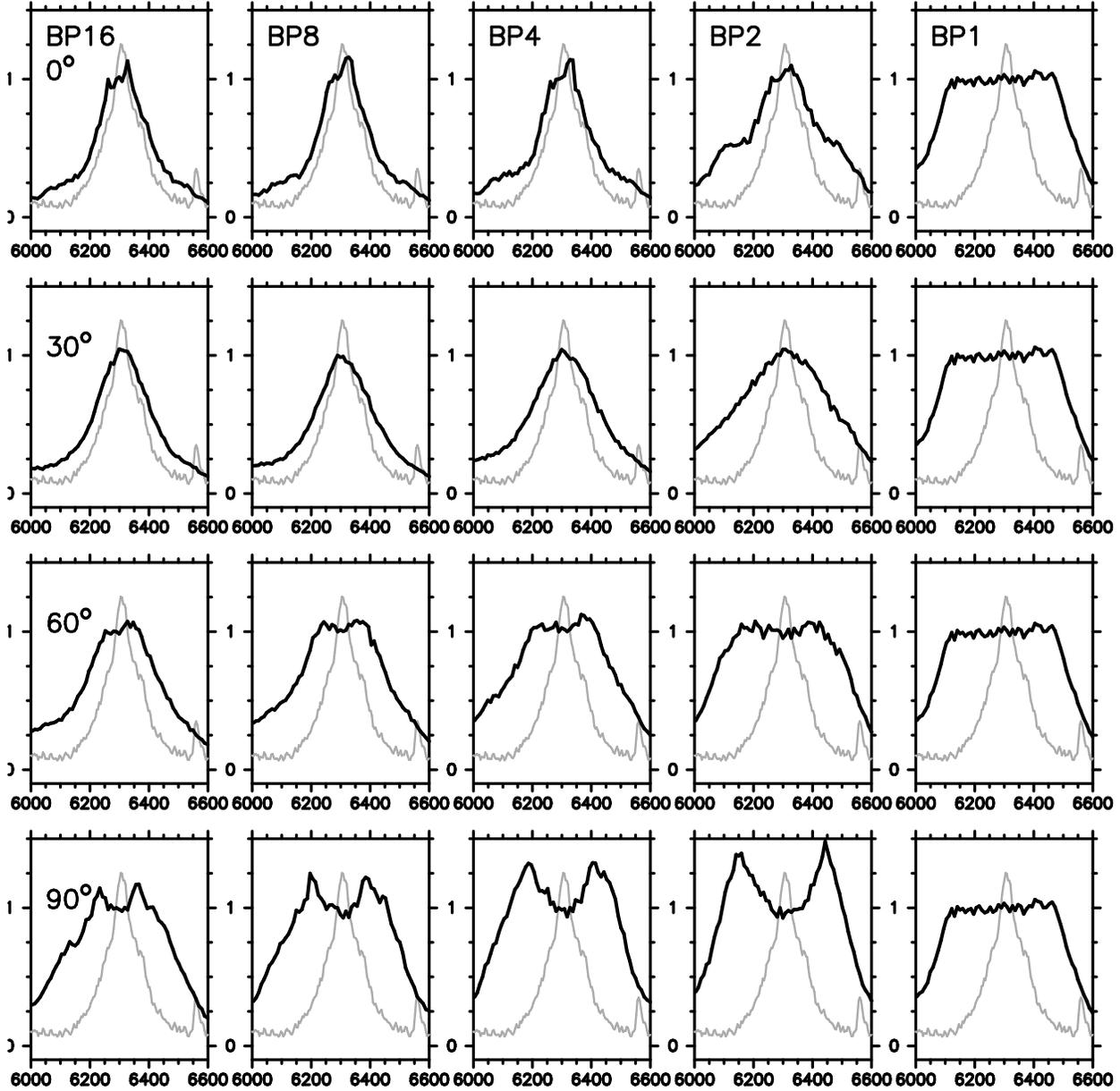}
\end{center}
\caption[] 
{The synthetic spectra of the 2D models 
around 6300\AA\ (black lines) 
as compared with the observed one of SN 1998bw at 337 days 
after the B maximum (gray lines). 
The synthetic spectra are predominantly contributed by [OI] 
6300\AA\ and 6363\AA. 
The models are those with $f_V=1.6$. 
See the caption of Fig. 11 for the meanings of each panel. 
\label{fig10}}
\end{figure}

\clearpage
\begin{figure}
\begin{center}
		\epsscale{1.0}
		\plotone{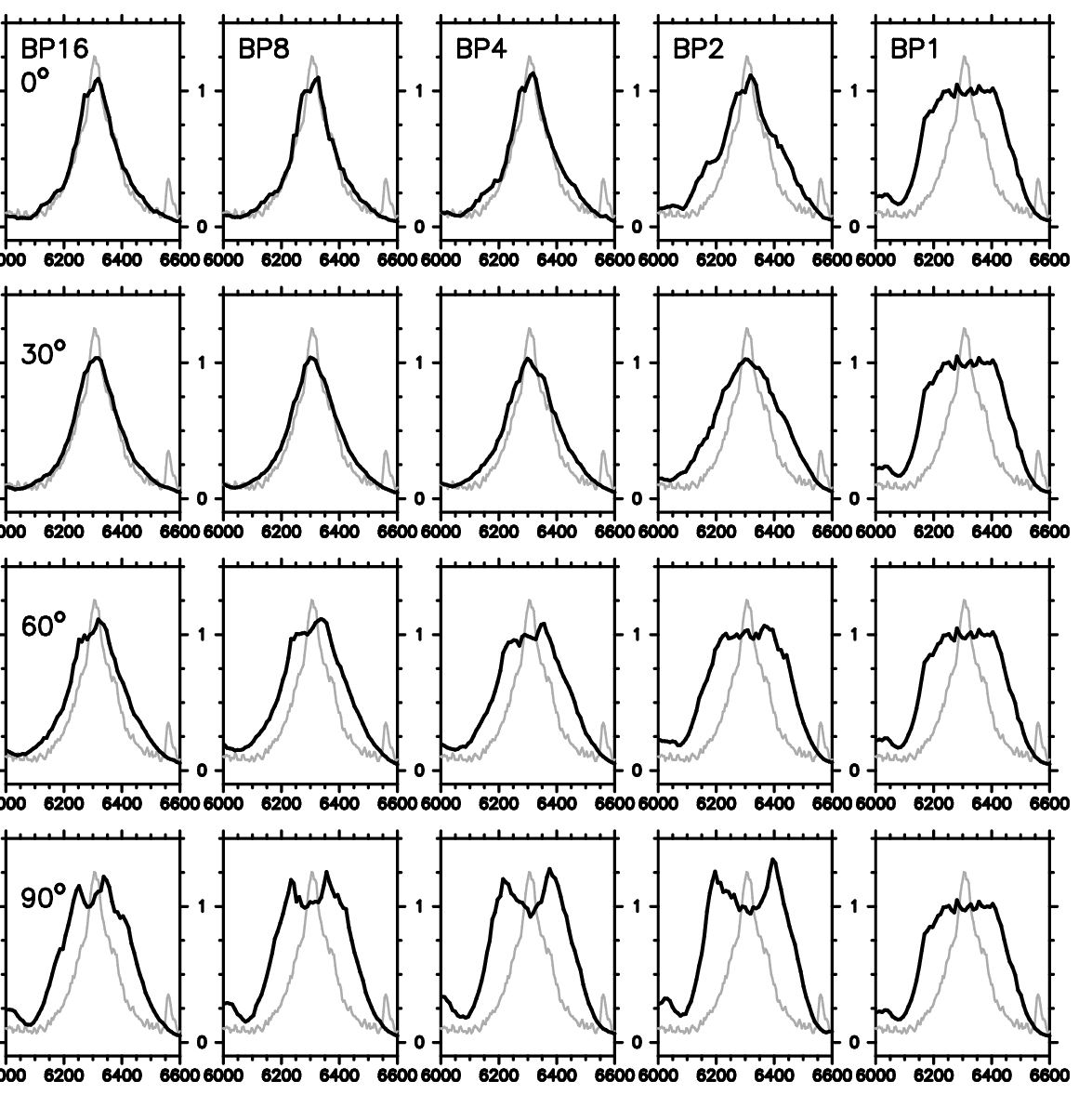}
\end{center}
\caption[] 
{The same with Fig. 14 ([OI] 6300\AA\ and 6363\AA), 
except the model parameter $f_V=1.15$. 
\label{fig11}}
\end{figure}

\clearpage
\begin{figure}
\begin{center}
		\epsscale{1.0}
		\plotone{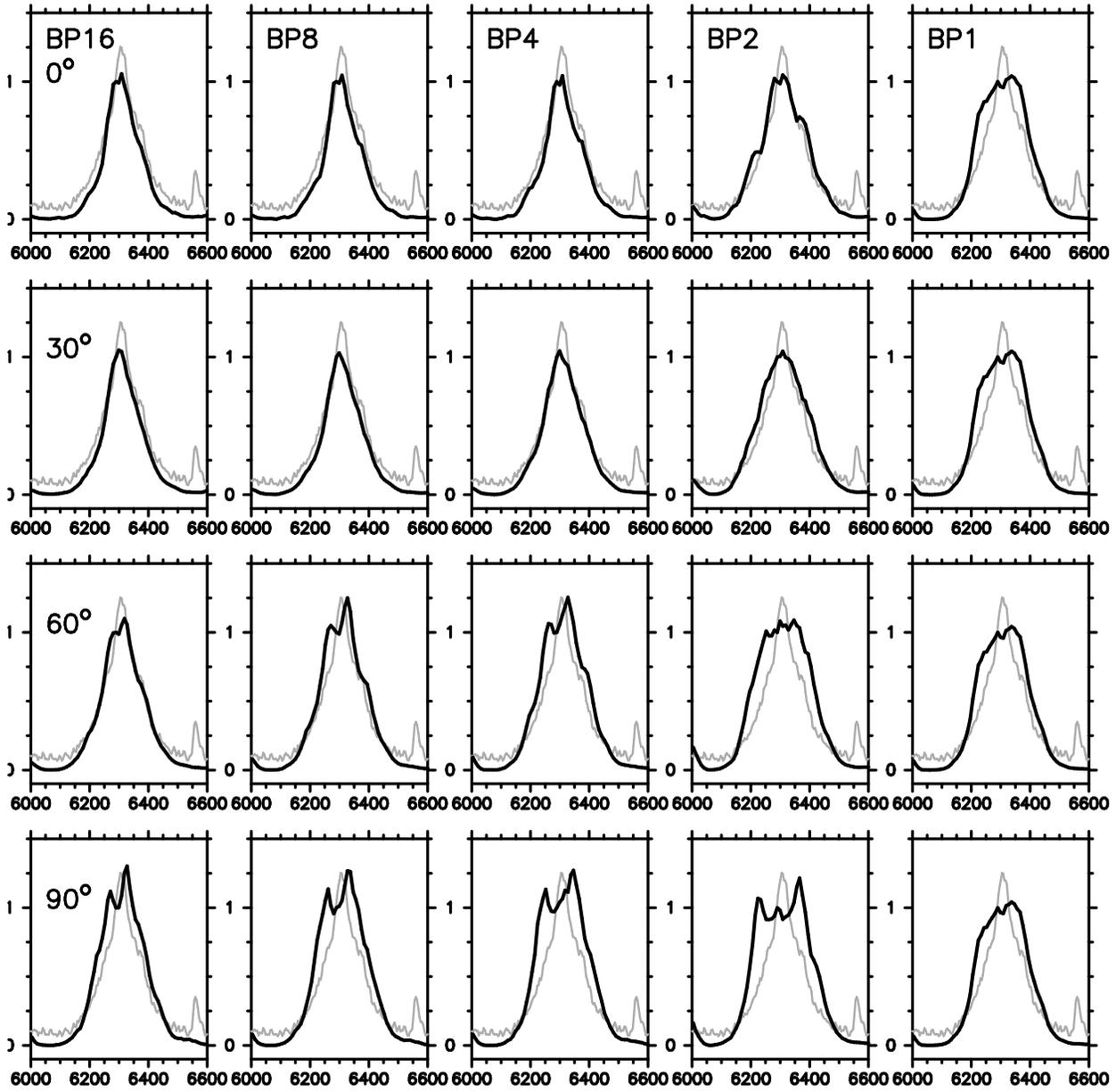}
\end{center}
\caption[] 
{The same with Fig. 14 ([OI] 6300\AA\ and 6363\AA), 
except the model parameter $f_V=0.7$. 
\label{fig12}}
\end{figure}

\clearpage
\begin{figure}
\begin{center}
		\epsscale{1.0}
		\plotone{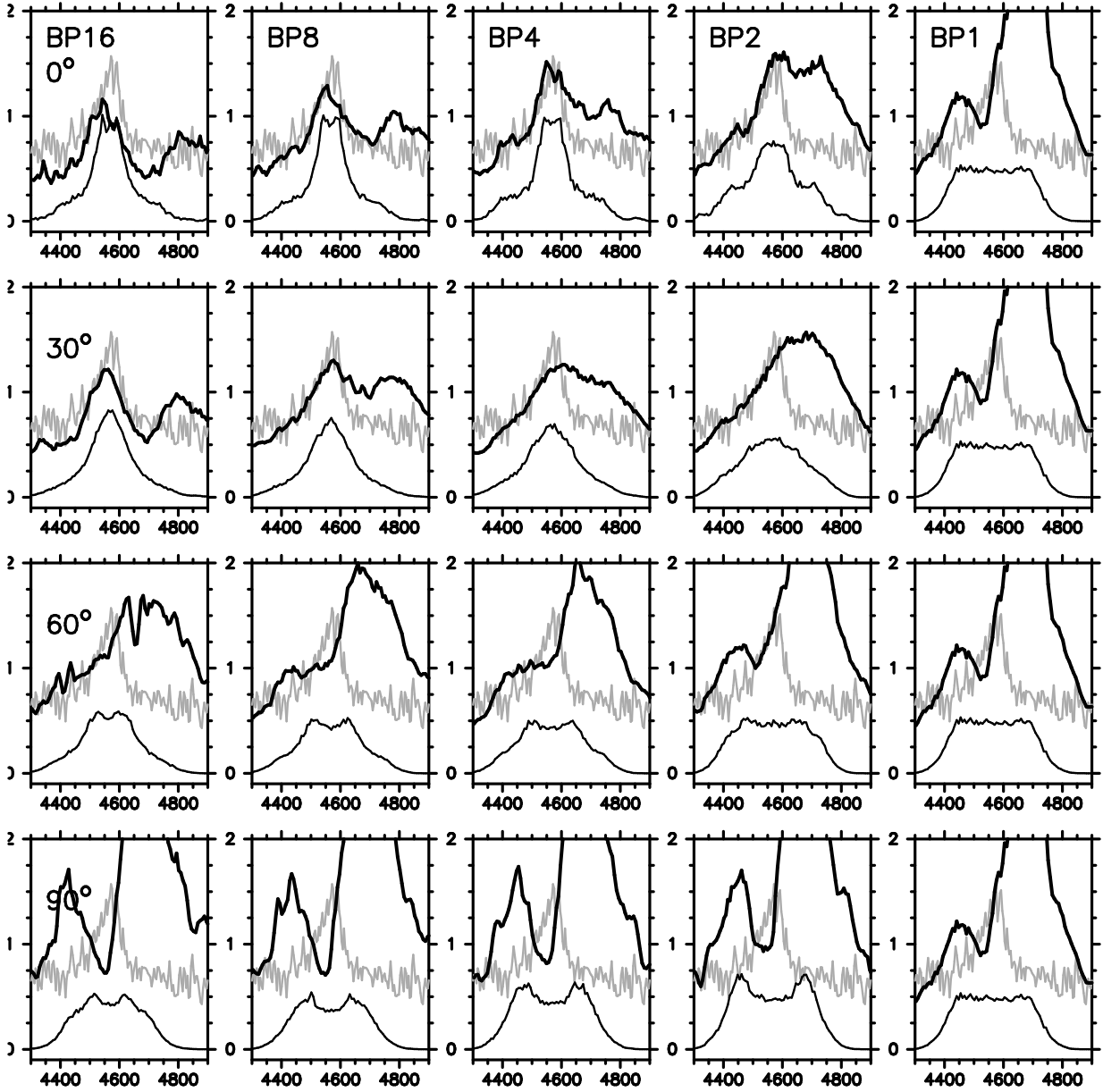}
\end{center}
\caption[] 
{The synthetic spectra of the 2D model 
around 4600\AA\ (black thick lines) 
as compared with the observed one of SN 1998bw at 337 days 
after the B maximum (gray lines). 
The synthetic spectra are mainly contributed by MgI] 
4570\AA\ and forest of [FeII] and [FeIII].  
The models are those with $f_V=1.6$. 
See the caption of Fig. 11 for the meanings of each panel. 
The contribution of MgI] 4570\AA\ is also shown 
(thin lines). 
\label{fig13}}
\end{figure}

\clearpage
\begin{figure}
\begin{center}
		\epsscale{1.0}
		\plotone{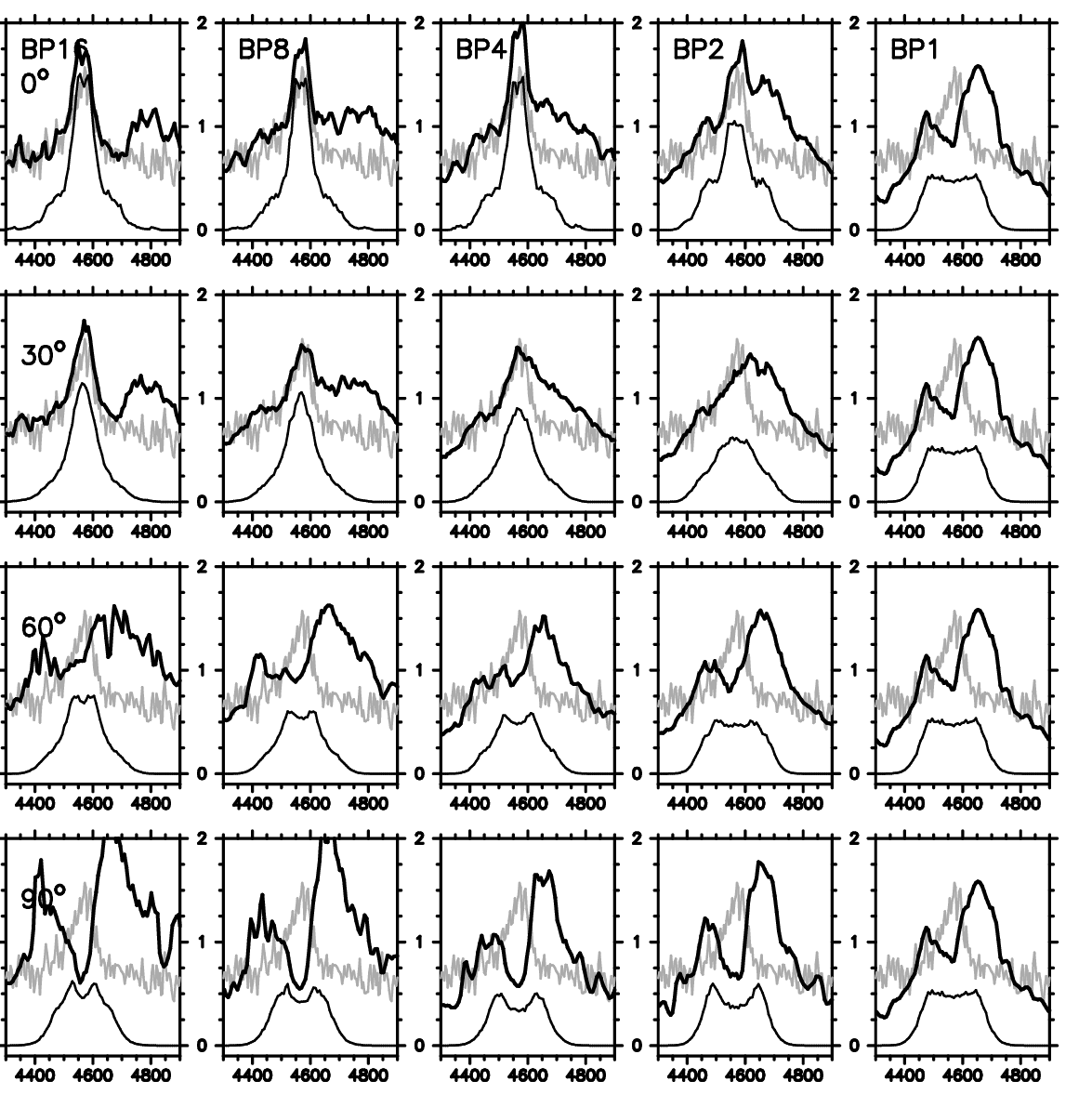}
\end{center}
\caption[] 
{The same with Fig. 17 (MgI] 4570\AA\ and [FeII], [FeIII]), 
except the model parameter $f_V=1.15$. 
\label{fig14}}
\end{figure}

\clearpage
\begin{figure}
\begin{center}
		\epsscale{1.0}
		\plotone{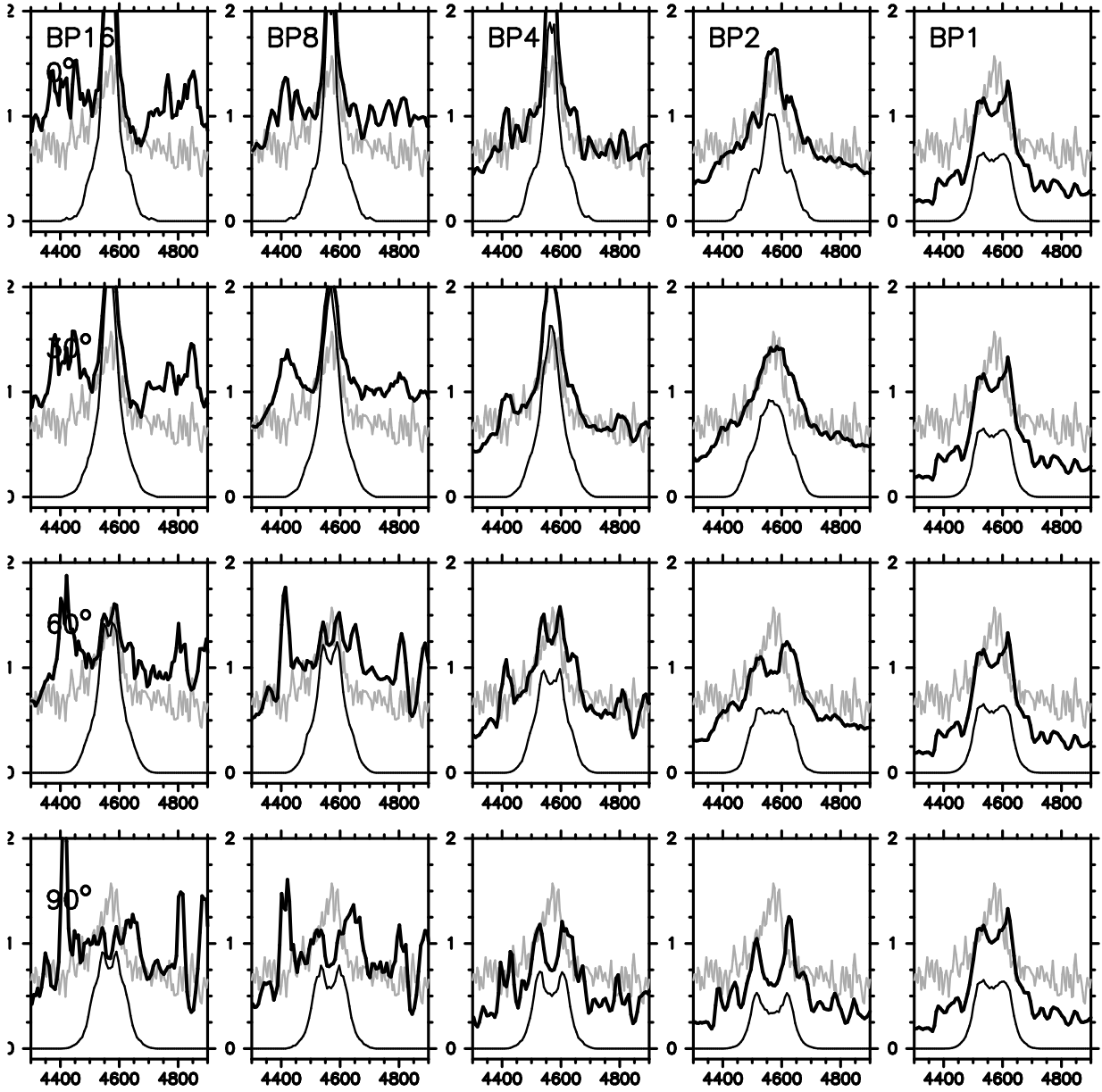}
\end{center}
\caption[] 
{The same with Fig. 17 (MgI] 4570\AA\ and [FeII], [FeIII]), 
except the model parameter $f_V=0.7$. 
\label{fig15}}
\end{figure}

\clearpage
\begin{figure}
\begin{center}
		\epsscale{1.0}
		\plotone{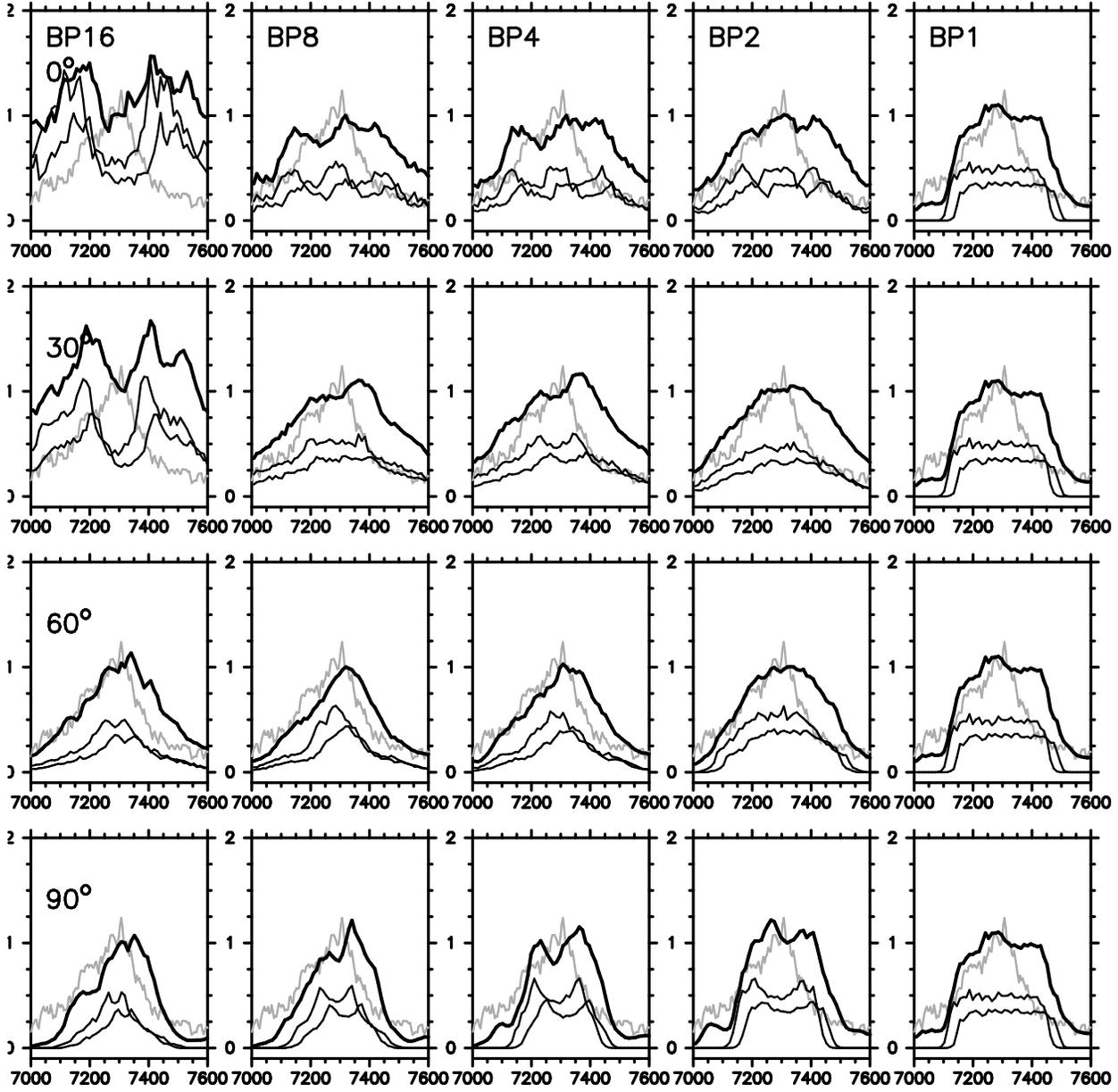}
\end{center}
\caption[] 
{The synthetic spectra of the 2D models around 7300\AA\ (black thick lines) 
as compared with the observed one of SN 1998bw at 337 days 
after the B maximum (gray lines). 
The synthetic spectra are mainly contributed by [CaII] 
7291\AA, 7324\AA, and [FeII] 7155\AA, 7172\AA, 7388\AA, and 7452\AA.  
The models are those with $f_V=1.6$. 
See the caption of Fig. 11 for the meanings of each panel. 
The contributions of [CaII] 7291\AA\ and 7324\AA\ are also shown (thin lines). 
\label{fig16}}
\end{figure}

\clearpage
\begin{figure}
\begin{center}
		\epsscale{1.0}
		\plotone{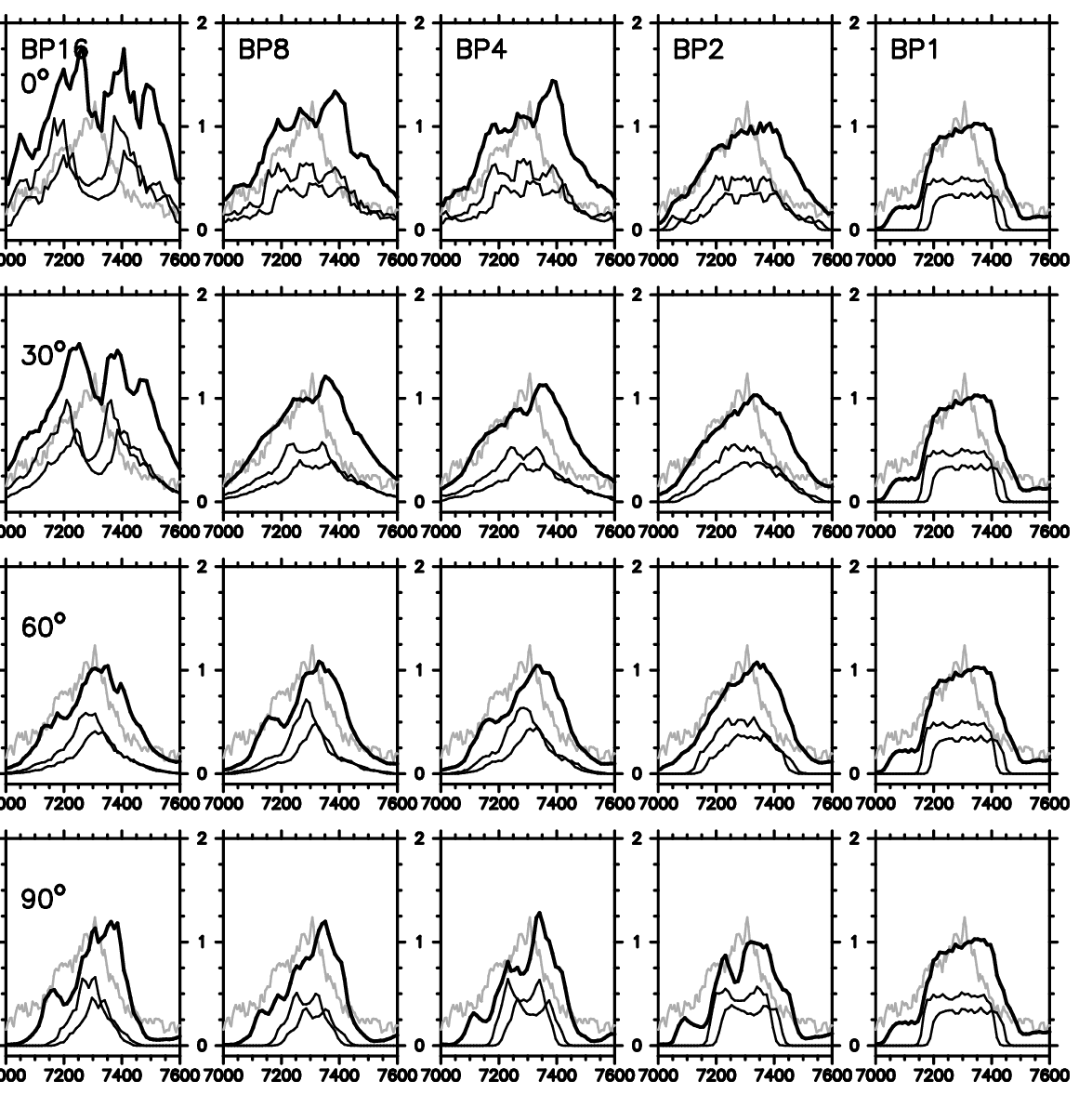}
\end{center}
\caption[] 
{The same with Fig. 20 ([CaII] 7291\AA, 7323\AA\, and [FeII]), 
except the model parameter $f_V=1.15$. 
\label{fig17}}
\end{figure}

\clearpage
\begin{figure}
\begin{center}
		\epsscale{1.0}
		\plotone{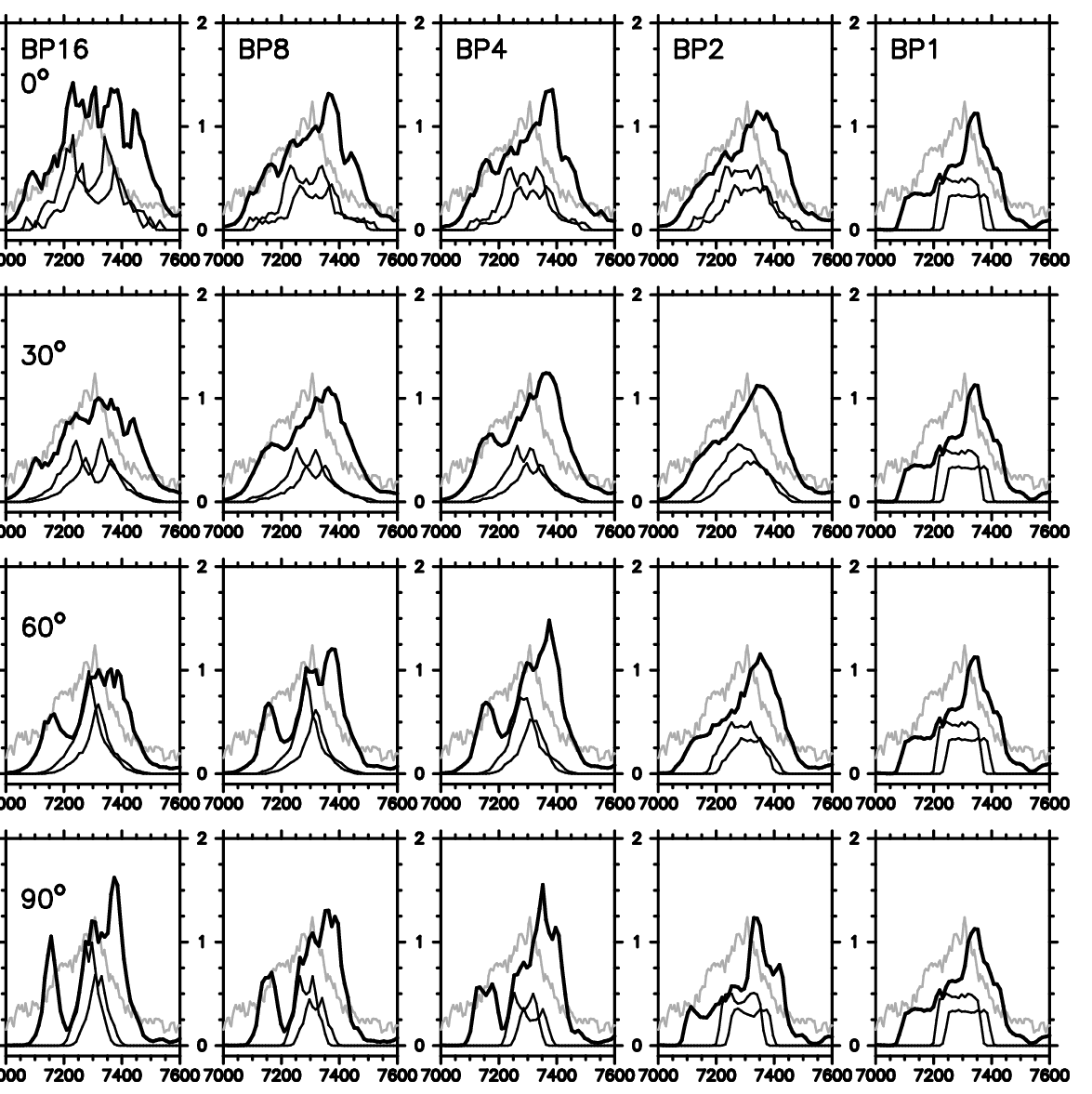}
\end{center}
\caption[] 
{The same with Fig. 20 ([CaII] 7291\AA, 7323\AA\, and [FeII]), 
except the model parameter $f_V=0.7$. 
\label{fig18}}
\end{figure}

\end{document}